%% file: root.tex
\documentclass[letterpaper, 10 pt, conference]{ieeeconf}  
\IEEEoverridecommandlockouts
\input{packages}

\begin{document}

\title{ \bf  Maximizing Safety and Efficiency for Cooperative Lane-Changing:\\ A Minimally Disruptive Approach}

\author{Andres S. Chavez Armijos, Anni Li, Christos G. Cassandras
\thanks{A. S. Chavez Armijos, A. Li, and C. G. Cassandras are
with the Division of Systems Engineering and the Center for Information and
Systems Engineering, Boston University, Brookline, MA 02446
(email:\{aschavez; anlianni; cgc\}@bu.edu).}
}

\maketitle
\begin{abstract}
\subfile{sections/abstract}
\end{abstract}

\section{Introduction}
\subfile{sections/introduction}
\section{Preliminaries}\label{sec:Preliminaries}
\subfile{sections/preliminaries}

\section{Problem Formulation}\label{sec:ProblemFormulation}
\subfile{sections/problem_formulation}
\section{From Planning to Execution}\label{sec:Planning2Execution}
\subfile{sections/planning2execution}

\section{Simulation Results}\label{sec:Simulation}
\subfile{sections/simulation_results}

\section{Conclusions and Future Work}\label{sec:Conclusions}
\subfile{sections/Conclusions}


\bibliographystyle{IEEEtran}
\begin{tiny}
\bibliography{bibliography}
\end{tiny}

\end{document}

%% file: packages.tex
\usepackage[OT1]{fontenc} 
\usepackage{cite}
\usepackage{amsmath,amssymb,amsfonts}
\usepackage[table,xcdraw]{xcolor}
\usepackage{textcomp}
\usepackage{blindtext}
\usepackage{subfiles} 
\usepackage{mathtools, nccmath}
\usepackage{physics}
\usepackage{tikz}
\usepackage{mathdots}
\usepackage{yhmath}
\usepackage{cancel}
\usepackage{color}
\usepackage{siunitx}
\usepackage{array}
\usepackage{multirow}
\usepackage{amssymb}
\usepackage{gensymb}
\usepackage{adjustbox}
\usepackage{tabularx}
\usepackage{extarrows}
\usepackage{booktabs}
\usepackage{diagbox}
\usepackage{hyperref}
\usepackage{verbatim}
\usepackage[normalem]{ulem}
\usepackage{subcaption}
\hypersetup{hidelinks, allcolors=black}
\useunder{\uline}{\ul}{}
\useunder{\uline}{\ul}{}
\usetikzlibrary{fadings}
\usetikzlibrary{patterns}
\usetikzlibrary{shadows.blur}
\usetikzlibrary{shapes}
\graphicspath{ {./Figures/} }

\DeclareMathOperator*{\argmin}{argmin}

\newtheorem{thm}{Theorem}[section]

\newtheorem{defn}{Definition}[section]

\newtheorem{rem}{Remark}

\newtheorem{assumption}{Assumption}
\newtheorem{prob}{Problem}

%% file: sections/abstract.tex
This paper addresses cooperative lane-changing maneuvers in mixed traffic, aiming to minimize traffic flow disruptions while accounting for uncooperative vehicles. The proposed approach adopts controllers combining 
Optimal control with Control Barrier Functions (OCBF controllers) which guarantee spatio-temporal constraints through the use of fixed-time convergence. Additionally, we introduce robustness  to disturbances by deriving a method for handling worst-case disturbances using the dual of a linear programming problem. We present a near-optimal solution that ensures safety, optimality, and robustness to changing behavior of uncooperative vehicles. Simulations demonstrate the effectiveness of the proposed approach in enhancing efficiency and safety.

%% file: sections/introduction.tex
Cooperative autonomous driving technology has received significant attention in recent years due to advancements in communication and sensor technologies. These advancements facilitate efficient and accurate information transmission between Connected Autonomous Vehicles (CAVs) and the surrounding environment. CAVs have the potential to greatly enhance traffic efficiency and safety in a variety of challenging traffic settings through effective trajectory planning. 

Among the critical aspects of cooperative highway driving, the automation of lane-changing maneuvers has increasingly become a focal point of research \cite{fisac2019hierarchical, chen2020cooperative, wang2021dynamic}.
However, it is crucial to minimize the overall negative impact, specifically traffic flow disruption, that cooperative maneuvers can cause. \emph{Disruption} refers to the cumulative effect of deceleration during maneuver executions on traffic flow. Previous studies, e.g.,  \cite{chen2022hierarchical},\cite{chalaki2023minimally},\cite{armijos2022cooperative}, have focused on selecting the optimal cooperative set and merging gap for a vehicle while maximizing efficiency through optimal control. In this context, efficiency encompasses the minimization of disruption, energy consumption, and time. However, this often assumes a constant (time-invariant) behavior of uncooperative vehicles and disregards possible disturbances, leading to unsafe policies during maneuver execution.

To ensure the safety of cooperative maneuvers, approximate solutions using Control Barrier Functions (CBFs) \cite{ames2016control},\cite{xiao2023safe} have been employed in \cite{xiao2021decentralized} and \cite{lyu2022adaptive}. In the context of automated lane-changing maneuvers, \cite{he2021rule} proposed a rule-based lane-changing strategy without cooperation using CBFs. However, these approaches overlook the optimality of cooperative maneuvers due to the conservative nature of CBFs, resulting in reduced maneuver efficiency and high traffic disruptions.

To address the trade-off between safety and optimality, \cite{xiao2021bridging} introduced the concept of Optimal control with Control Barrier Functions (OCBFs), utilizing an analytically obtained unconstrained solution to an Optimal Control Problem (OCP) as a reference trajectory to track subject to CBF-based constraints, thus bridging the gap between the \emph{safety guarantees} of CBFs and the \emph{optimality} of OCP solutions. This approach has been applied in the context of cooperative driving in conflict settings such as merging \cite{xiao2021decentralized} and roundabouts \cite{xu2021decentralized}. However, for cooperative lane-changing scenarios, ensuring the timely achievement of terminal conditions is crucial for the feasibility and efficiency of the maneuvers. The presence of such temporal constraints motivates this work.

In this paper, we consider cooperative lane-changing maneuvers in mixed traffic, as depicted in Fig. \ref{fig:cav_set_selection}, aiming to perform minimally disruptive maneuvers despite the presence of uncooperative vehicles. Specifically, our contributions are threefold: First, we propose a novel method that combines OCBFs with spatio-temporal constraints through the use of fixed-time convergence (FxT-OCBF) for cooperative lane-changing maneuvers. Second, we introduce a method that computes the dual of a linear programming problem to handle additive disturbances. Third, building upon \cite{garg2022fixed} and \cite{xiao2021bridging}, we extend previous works in \cite{armijos2022cooperative} and \cite{chalaki2023minimally} to provide a near-optimal solution for the minimally disruptive cooperative lane-changing problem. The proposed method demonstrates robustness to changes in the behavior of uncooperative vehicles while ensuring safety and near-optimality. 

The remainder of this paper is structured as follows. Section \ref{sec:Preliminaries} briefly discusses the theory behind CBFs and fixed-time convergence. Section \ref{sec:ProblemFormulation} summarizes our previous work, followed by the formulation of the lane changing problem. Our current results are introduced in Section \ref{sec:Planning2Execution}, and we conduct simulations in Section \ref{sec:Simulation} to demonstrate the efficacy of our approach. Finally, we provide concluding remarks together with future work in Section \ref{sec:Conclusions}.

%% file: sections/preliminaries.tex
In this paper, we address an optimal control problem (OCP) for a nonlinear control affine system given by:
\begin{equation}
\dot{x}=f(x(t))+g(x(t))u+d(t,x),
\label{eq:affine_system}
\end{equation}
Here, $x\in\mathbb{R}^n$ and $u\in\mathcal{U}\subset\mathbb{R}^m$ represent the state and control input vectors, respectively. The control constraint set $\mathcal{U}:=\left(u \in \mathbb{R}^n: u_{\min } \leq u \leq u_{\max }\right)$ defines the allowable control inputs. The functions $f:\mathbb{R}^n\rightarrow\mathbb{R}^n$ and $g:\mathbb{R}^m\rightarrow\mathbb{R}^{n\times m}$ are continuous mappings, while $d:\mathbb{R}^n\rightarrow\mathbb{R}^n$ represents an additive disturbance accounting for modeling errors or environmental disturbances. 
\begin{assumption}
There exists a positive constant $\gamma$ such that for all $t\geq 0$ and $x\in D\subset\mathbb{R}^n$, the disturbance $d$ is bounded by $|d(t,x)|\leq\gamma$.
\end{assumption}

We use Control Barrier Functions (CBFs) and Control Lyapunov Functions (CLFs) to map the safety constraints and goal constraints (see \cite{xiao2023safe} for details). Our objective is to ensure that system \eqref{eq:affine_system} remains within a safety set $S_S:=\{x|h_S(x)\geq0\}$. This guarantees the system's safety under closed-loop dynamics. Simultaneously, we aim to guide the closed-loop trajectories of \eqref{eq:affine_system} towards a goal set $S_G:=\{x|h_G(x)\leq0\}$ within a specified time duration $T_p>0$. It is important to note that $h_S(\boldsymbol{x})\cap h_G(\boldsymbol{x})\neq\emptyset$ with $h_S(\boldsymbol{x})$ and $h_G(\boldsymbol{x})$ representing functions that characterize the sets $S_S$ and $S_G$, respectively.

\begin{defn}(\textbf{Fixed-Time Domain of Attraction \cite{garg2022fixed}}) 
A compact set $S$ is said to be a Fixed-Time Domain of Attraction (FxT-DoA) with time $T>0$ if it is completely contained within a set $D$ in $\mathbb{R}^n$, and the following conditions hold true for the closed-loop system \eqref{eq:affine_system} under input $u$:
i) For all initial states $x(0)\in D$, the solution $x(t)$ remains in $D$ for all time $t$ between 0 and $T$. 
ii) There exists a time $T_0$ between 0 and $T$ such that the limit of $x(t)$ as $t$ approaches $T_0$ and belongs to $S$.
\end{defn}

\begin{thm}(\textbf{FxT-CLF-CBF-$\mathbf{S}_\mathbf{G}$} \cite{garg2019control})\label{thm:conditions_safety_convergence}
For a given user-specified $T_{ud}$, if there exist parameters $\alpha_{i1}$, $\alpha_{i2}$, $\gamma_{i1}>1$, and $0<\gamma_{i2}<1$ for $i\in\Sigma$ such that
$
T_{ud} \geq \max_{i\in\Sigma}\lbrack\frac{1}{\alpha_{i 1}\left(\gamma_{i 1}-1\right)}+\frac{1}{\alpha_{i 2}\left(1-\gamma_{i 2}\right)}\rbrack,
$
and a control input $u(t)$ satisfies
\begin{subequations}
     \begin{align}
        & \inf _{u \in \mathcal{U}}\left\{\mathcal{L}_f h_S(x)+\mathcal{L}_g h_S(x) u\right\}+\beta_s\left(h_S(x)\right)\geq 0 \label{sub_eq:cbf_constraint}\\
        & \inf _{u \in \mathcal{U}}\left\{\mathcal{L}_f h_{G}(x)+\mathcal{L}_g h_{G}(x) u \right\}+ \alpha_{1} \max \left\{0, h_{G}(x)\right\}^{\gamma_{1}} \notag \\
        &\quad \quad -\alpha_{2} \max \left\{0, h_{G}(x)\right\}^{\gamma_{2}}\leq 0, \label{sub_eq:fxt_clf_constraint}
    \end{align}
\label{eq:conditions_safety_convergence}
\end{subequations}
where $\beta_s(\cdot)$ denotes an extended class-$\mathcal{K}$ function, then under the control input $u$, the closed-loop trajectories satisfy $x(t)\in S_S$ and $x(t)\in S_G$ for all $x\in\mathbb{R}^n\setminus S$, where the convergence time $T$ satisfies $T\leq \frac{\mu\pi}{2\sqrt{\alpha_1\alpha_2}}\leq T_{ud}$. 
\end{thm}

In practical terms, for an OCP with quadratic objectives, we divide the time interval $[0, T]$ into equal-sized steps of duration $\Delta t$. Within each step, indexed by $k$, the state is assumed to remain constant at its initial value, and we also assume a constant control input. To satisfy Theorem \ref{thm:conditions_safety_convergence}, we formulate a quadratic programming (QP) problem that computes a control $u_k$ satisfying \eqref{eq:conditions_safety_convergence}. Considering the vector $z=\left[u_k^\top ; \delta_g; \delta_s\right]^\top$, we solve the following optimization problem at each time step $k$:
\begin{subequations}
    \begin{gather}
    \min \limits_{u_k\in U, \delta_g, \delta_s\geq0} \frac{1}{2} z^\top H z+F^\top z \\
    \text { s.t.} \quad
    \mathcal{L}_f h_G(x)+\mathcal{L}_g h_G(x) u \leq \delta_g h_G(x)- \notag \\
        \alpha_1 \max \left\{0, h_G(x)\right\}^{\gamma_1} -\alpha_2 \max \left\{0, h_G(x)\right\}^{\gamma_2} \\
    \mathcal{L}_f h_S(x)+\mathcal{L}_g h_S(x) u +\delta_s h_S(x)\geq 0
    \end{gather}
    \label{eq:qp_cbf_clf_fxt}
\end{subequations}
Here, $\delta_g$ and $\delta_s$ are the gains of linear extended class-$\mathcal{K}$ functions. 
Matrices $H$ and $F$ are suitably chosen, with the term $F^Tz$ penalizing positive values of $\delta_1$. Additionally, fixed parameters $\alpha_1$, $\alpha_2$, $\gamma_1$, and $\gamma_2$ are chosen as $\alpha_1=\alpha_2=\frac{\mu \pi}{2 T_{ud}}$, $\gamma_1=1+\frac{1}{\mu}$, and $\gamma_2=1-\frac{1}{\mu}$, where $\mu>1$.
\begin{rem}
Note that in \eqref{eq:qp_cbf_clf_fxt}, $\delta_g$ and $\delta_s$ are optimization variables that relax the enforcement of the FxT-CLF-CBF-$S_G$ conditions. Including the constraint $\delta_s\geq 0$ guarantees safety but can void fixed-time convergence in some cases. In this paper, we adopt this convention to prioritize safety over efficiency in the trajectory of the lane-changing maneuvers.
\end{rem}
\vspace*{-2mm}

%% file: sections/problem_formulation.tex
\begin{figure}[ptb]
    \centering
    \vspace*{1mm}
    \begin{adjustbox}{width=\linewidth, height = 2.5cm,center}
\tikzset{every picture/.style={line width=0.75pt}} 

\begin{tikzpicture}[x=0.75pt,y=0.75pt,yscale=-1,xscale=1]
\draw [line width=3]    (116.81,39.52) -- (305.1,39.21) -- (511.37,39.35) ;
\draw [line width=3]    (116.43,136.42) -- (284.61,137.02) -- (510.99,136.59) ;
\draw [color={rgb, 255:red, 248; green, 231; blue, 28 }  ,draw opacity=1 ][fill={rgb, 255:red, 248; green, 231; blue, 28 }  ,fill opacity=1 ][line width=3]  [dash pattern={on 11.25pt off 9.75pt}]  (117.33,90.28) -- (512,89.77) ;
\draw (382.84,109.67) node [rotate=-0.02,xslant=0] {\includegraphics[width=26.23pt,height=16.6pt]{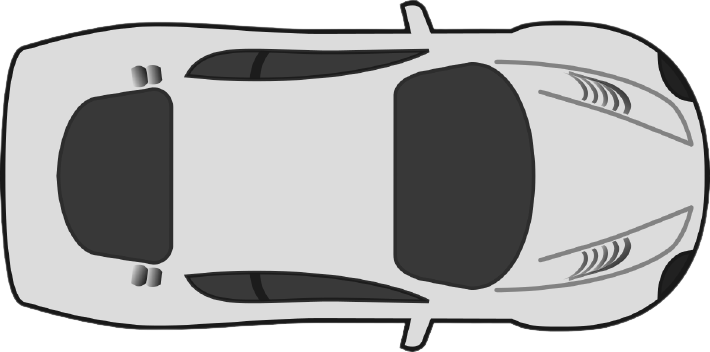}};
\draw (323.48,63.06) node [rotate=-0.02,xslant=0] {\includegraphics[width=26.23pt,height=16.6pt]{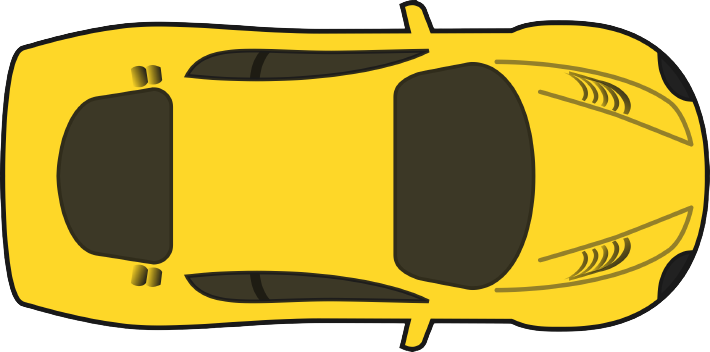}};
\draw (400.04,63.1) node [rotate=-0.02,xslant=0] {\includegraphics[width=26.23pt,height=16.6pt]{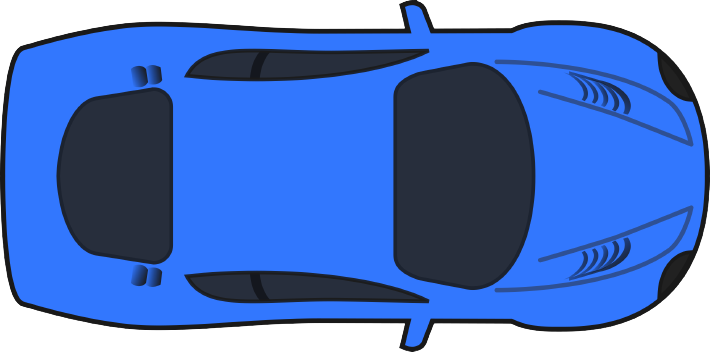}};
\draw (299.03,109.17) node [rotate=-0.02,xslant=0] {\includegraphics[width=26.23pt,height=16.6pt]{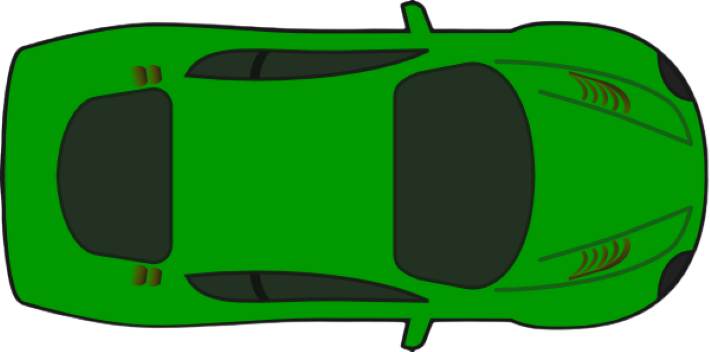}};
\draw (253.3,62.57) node [rotate=-0.02,xslant=0] {\includegraphics[width=26.23pt,height=16.6pt]{car_yellow.png}};
\draw (183.84,63.02) node [rotate=-0.02,xslant=0] {\includegraphics[width=26.23pt,height=16.6pt]{car_blue.png}};
\draw (195.21,110.55) node [rotate=-0.02,xslant=0] {\includegraphics[width=26.23pt,height=16.6pt]{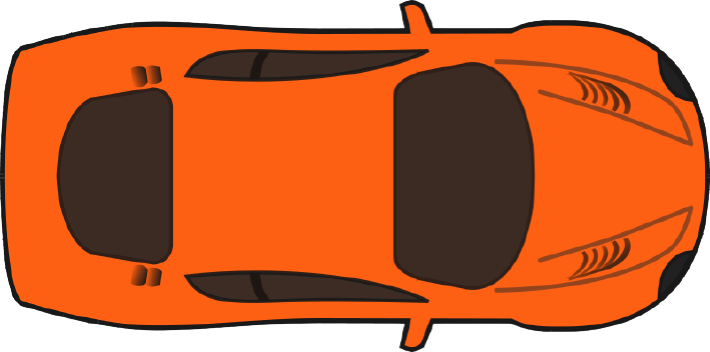}};
\draw  [color={rgb, 255:red, 208; green, 2; blue, 27 }  ,draw opacity=1 ][dash pattern={on 4.5pt off 4.5pt}] (150.4,55.5) .. controls (150.41,50.8) and (154.22,46.99) .. (158.92,46.99) -- (425.18,47.09) .. controls (429.88,47.09) and (433.69,50.9) .. (433.68,55.6) -- (433.66,81.11) .. controls (433.65,85.81) and (429.84,89.62) .. (425.14,89.62) -- (158.87,89.53) .. controls (154.18,89.52) and (150.37,85.71) .. (150.38,81.02) -- cycle ;
\draw  [dash pattern={on 0.75pt off 0.75pt}]  (317.37,108.85) .. controls (319.06,107.2) and (320.72,107.22) .. (322.37,108.91) -- (326.48,108.97) -- (334.48,109.07) ;
\draw [shift={(336.48,109.09)}, rotate = 180.73] [color={rgb, 255:red, 0; green, 0; blue, 0 }  ][line width=0.75]    (10.93,-3.29) .. controls (6.95,-1.4) and (3.31,-0.3) .. (0,0) .. controls (3.31,0.3) and (6.95,1.4) .. (10.93,3.29)   ;
\draw  [color={rgb, 255:red, 245; green, 166; blue, 35 }  ,draw opacity=1 ][dash pattern={on 4.5pt off 4.5pt}] (223.09,56.4) .. controls (223.1,52.16) and (226.53,48.73) .. (230.77,48.73) -- (342.01,48.77) .. controls (346.24,48.77) and (349.68,52.21) .. (349.67,56.44) -- (349.65,79.45) .. controls (349.64,83.69) and (346.2,87.12) .. (341.97,87.12) -- (230.73,87.08) .. controls (226.49,87.08) and (223.06,83.64) .. (223.07,79.41) -- cycle ;
\draw [color={rgb, 255:red, 245; green, 166; blue, 35 }  ,draw opacity=1 ]   (315.44,44.67) -- (315.88,28.7) ;
\draw [shift={(315.93,26.7)}, rotate = 91.57] [color={rgb, 255:red, 245; green, 166; blue, 35 }  ,draw opacity=1 ][line width=0.75]    (10.93,-3.29) .. controls (6.95,-1.4) and (3.31,-0.3) .. (0,0) .. controls (3.31,0.3) and (6.95,1.4) .. (10.93,3.29)   ;
\draw [color={rgb, 255:red, 208; green, 2; blue, 27 }  ,draw opacity=1 ]   (407.8,43.46) -- (407.43,27.95) ;
\draw [shift={(407.38,25.95)}, rotate = 88.61] [color={rgb, 255:red, 208; green, 2; blue, 27 }  ,draw opacity=1 ][line width=0.75]    (10.93,-3.29) .. controls (6.95,-1.4) and (3.31,-0.3) .. (0,0) .. controls (3.31,0.3) and (6.95,1.4) .. (10.93,3.29)   ;
\draw (460.84,62.67) node [rotate=-0.02,xslant=0] {\includegraphics[width=26.23pt,height=16.6pt]{car_white.png}};

\draw (314.98,108.48) node [anchor=north west][inner sep=0.75pt]  [rotate=-0.04] [align=left] {$v_C(t) $};
\draw (374.79,117.53) node [anchor=north west][inner sep=0.75pt]  [font=\small] [align=left] {$\displaystyle U$};
\draw (285.07,118.71) node [anchor=north west][inner sep=0.75pt]  [font=\small] [align=left] {$\displaystyle C$};
\draw (317.8,70.56) node [anchor=north west][inner sep=0.75pt]  [font=\small] [align=left] {$\displaystyle \hat{2}$};
\draw (244.88,69.89) node [anchor=north west][inner sep=0.75pt]  [font=\small] [align=left] {$\displaystyle \hat{3}$};
\draw (395.19,69.41) node [anchor=north west][inner sep=0.75pt]  [font=\small] [align=left] {$\displaystyle \hat{1}$};
\draw (175.99,68.41) node [anchor=north west][inner sep=0.75pt]  [font=\small] [align=left] {$\displaystyle \hat{4}$};
\draw (390.26,11.19) node [anchor=north west][inner sep=0.75pt]  [color={rgb, 255:red, 208; green, 2; blue, 27 }  ,opacity=1 ] [align=left] {$\displaystyle \mathcal{S}(t_f)$};
\draw (280.86,8.41) node [anchor=north west][inner sep=0.75pt]  [color={rgb, 255:red, 245; green, 166; blue, 35 }  ,opacity=1 ] [align=left] {$\displaystyle \left( i^{*} ,i^{*} +1\right)$};
\draw (452.79,71.53) node [anchor=north west][inner sep=0.75pt]  [font=\small] [align=left] {$\displaystyle F$};

\end{tikzpicture}
    \end{adjustbox}
    \caption{\centering{Cooperative vehicle set $\mathcal{S}(t)$, and optimal merging slot $(i^*,i^*+1) \in \mathcal{S}(t)$ selection diagram}}
    \label{fig:cav_set_selection}
    \vspace*{-\baselineskip} \vspace*{-2mm}
\end{figure}

This section provides a review of our previous work on cooperative lane-changing maneuvers \cite{chalaki2023minimally, armijos2022cooperative}, which serves as the foundation for the optimal control solution proposed in this paper which includes temporal constraints. The maneuver comprises longitudinal and lateral segments, where the lateral maneuver is executed when there is sufficient space for merging. Fig. \ref{fig:cav_set_selection} illustrates the presence of two uncooperative vehicles, referred to as $U$ and $F$, traveling in the slow and fast lanes, respectively. Our focus is on a specific CAV, denoted as $C$, which performs an automated lane change initiated when it is necessary to overtake vehicle $U$. The maneuver begins when $C$ detects vehicle $U$ at a distance of $d_{\text{start}}$ ahead. Additionally, we consider a group of CAVs traveling behind vehicle $F$.

To describe cooperative behavior, we define the set $\mathcal{S}(t)=\{1,\ldots,N\}$, representing CAVs near $C$ capable of cooperation at time $t$. Indices 1 and $N$ correspond to the CAVs farthest ahead and farthest behind $C$, respectively. Each CAV $i\in\mathcal{S}(t)$ can sense the states of $U$, $F$, and other uncooperative vehicles. Due to space limitations, we omit details on lateral trajectory optimization found in \cite{chen2020cooperative}.

\begin{figure}[tpb]
    \centering
    \begin{adjustbox}{width=0.8\linewidth, height = 2.3cm, center}
        \tikzset{every picture/.style={line width=0.75pt}} 
        \begin{tikzpicture}[x=0.75pt,y=0.75pt,yscale=-1,xscale=1]
        
        \draw [color={rgb, 255:red, 248; green, 231; blue, 28 }  ,draw opacity=1 ][fill={rgb, 255:red, 248; green, 231; blue, 28 }  ,fill opacity=1 ][line width=3]  [dash pattern={on 22.5pt off 7.5pt}]  (57.43,138.11) -- (583.81,136.69) ;
        \draw  [fill={rgb, 255:red, 0; green, 0; blue, 0 }  ,fill opacity=1 ] (179.44,144.38) .. controls (179.2,143.53) and (179.7,142.65) .. (180.55,142.42) -- (203.8,135.97) .. controls (204.65,135.73) and (205.53,136.23) .. (205.76,137.08) -- (207.35,142.8) .. controls (207.59,143.65) and (207.09,144.53) .. (206.24,144.76) -- (182.99,151.22) .. controls (182.14,151.45) and (181.26,150.95) .. (181.03,150.1) -- cycle ;
        \draw  [fill={rgb, 255:red, 0; green, 0; blue, 0 }  ,fill opacity=1 ] (194.44,204.38) .. controls (194.2,203.53) and (194.7,202.65) .. (195.55,202.42) -- (218.8,195.97) .. controls (219.65,195.73) and (220.53,196.23) .. (220.76,197.08) -- (222.35,202.8) .. controls (222.59,203.65) and (222.09,204.53) .. (221.24,204.76) -- (197.99,211.22) .. controls (197.14,211.45) and (196.26,210.95) .. (196.03,210.1) -- cycle ;
        \draw  [fill={rgb, 255:red, 0; green, 0; blue, 0 }  ,fill opacity=1 ] (255.77,198.58) .. controls (255.23,197.88) and (255.36,196.88) .. (256.06,196.34) -- (275.16,181.61) .. controls (275.86,181.07) and (276.86,181.2) .. (277.4,181.9) -- (281.02,186.6) .. controls (281.56,187.3) and (281.43,188.3) .. (280.73,188.84) -- (261.63,203.57) .. controls (260.93,204.11) and (259.93,203.98) .. (259.39,203.28) -- cycle ;
        \draw  [fill={rgb, 255:red, 0; green, 0; blue, 0 }  ,fill opacity=1 ] (241.77,133.58) .. controls (241.23,132.88) and (241.36,131.88) .. (242.06,131.34) -- (261.16,116.61) .. controls (261.86,116.07) and (262.86,116.2) .. (263.4,116.9) -- (267.02,121.6) .. controls (267.56,122.3) and (267.43,123.3) .. (266.73,123.84) -- (247.63,138.57) .. controls (246.93,139.11) and (245.93,138.98) .. (245.39,138.28) -- cycle ;
        \draw (241.09,164.34) node [rotate=-348.35] {\includegraphics[width=91.22pt,height=57.34pt]{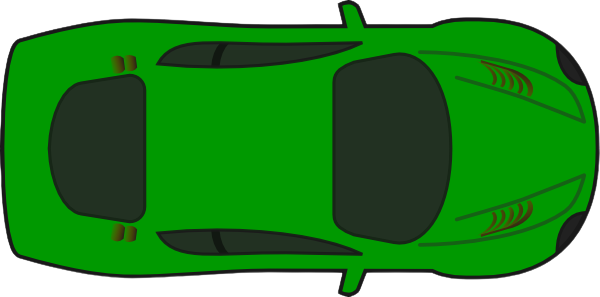}};
        \draw [line width=3]    (16,37.87) -- (334.76,35.7) -- (583.81,35.7) ;
        \draw [line width=3]    (16,238.87) -- (297.62,239.03) -- (583.81,239.03) ;
        \draw (438.53,187.25) node  {\includegraphics[width=91.99pt,height=53.37pt]{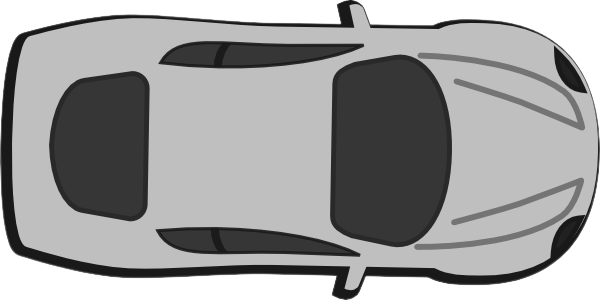}};
        \draw (461.97,83.76) node  {\includegraphics[width=129.04pt,height=112.85pt]{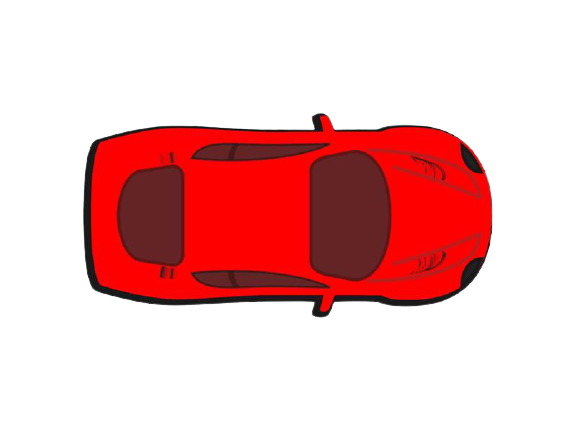}};
        \draw (92.73,83.5) node  {\includegraphics[width=131.6pt,height=115.95pt]{redVehicleTopView.png}};
        \draw [color={rgb, 255:red, 208; green, 2; blue, 27 }  ,draw opacity=1 ][fill={rgb, 255:red, 208; green, 2; blue, 27 }  ,fill opacity=1 ] [dash pattern={on 3.75pt off 3pt on 7.5pt off 1.5pt}]  (42.82,83.15) -- (9.38,83.15) ;
        \draw [shift={(7.38,83.15)}, rotate = 360] [color={rgb, 255:red, 208; green, 2; blue, 27 }  ,draw opacity=1 ][line width=0.75]    (10.93,-3.29) .. controls (6.95,-1.4) and (3.31,-0.3) .. (0,0) .. controls (3.31,0.3) and (6.95,1.4) .. (10.93,3.29)   ;
        \draw [color={rgb, 255:red, 208; green, 2; blue, 27 }  ,draw opacity=1 ] [dash pattern={on 3.75pt off 3pt on 7.5pt off 1.5pt}]  (522.12,82.77) -- (553.51,82.77) ;
        \draw [shift={(555.51,82.77)}, rotate = 180] [color={rgb, 255:red, 208; green, 2; blue, 27 }  ,draw opacity=1 ][line width=0.75]    (10.93,-3.29) .. controls (6.95,-1.4) and (3.31,-0.3) .. (0,0) .. controls (3.31,0.3) and (6.95,1.4) .. (10.93,3.29)   ;
        \draw [color={rgb, 255:red, 208; green, 2; blue, 27 }  ,draw opacity=1 ][line width=2.25]  [dash pattern={on 6.75pt off 4.5pt}]  (131.2,187.03) -- (325.2,187.03) ;
        \draw [color={rgb, 255:red, 208; green, 2; blue, 27 }  ,draw opacity=1 ][line width=2.25]  [dash pattern={on 6.75pt off 4.5pt}]  (131.2,187.03) -- (332.28,145.84) ;
        \draw [shift={(336.2,145.03)}, rotate = 168.42] [color={rgb, 255:red, 208; green, 2; blue, 27 }  ,draw opacity=1 ][line width=2.25]    (17.49,-5.26) .. controls (11.12,-2.23) and (5.29,-0.48) .. (0,0) .. controls (5.29,0.48) and (11.12,2.23) .. (17.49,5.26)   ;
        \draw [color={rgb, 255:red, 65; green, 117; blue, 5 }  ,draw opacity=1 ][line width=1.5]    (297.2,151.03) -- (325.09,117.18) ;
        \draw [shift={(327,114.87)}, rotate = 129.49] [color={rgb, 255:red, 65; green, 117; blue, 5 }  ,draw opacity=1 ][line width=1.5]    (14.21,-4.28) .. controls (9.04,-1.82) and (4.3,-0.39) .. (0,0) .. controls (4.3,0.39) and (9.04,1.82) .. (14.21,4.28)   ;
        \draw  [draw opacity=0] (304.69,141.21) .. controls (306.37,142.27) and (307.48,144.07) .. (307.49,146.13) .. controls (307.5,147.92) and (306.67,149.52) .. (305.35,150.61) -- (301.25,146.15) -- cycle ; \draw   (304.69,141.21) .. controls (306.37,142.27) and (307.48,144.07) .. (307.49,146.13) .. controls (307.5,147.92) and (306.67,149.52) .. (305.35,150.61) ;  
        \draw  [draw opacity=0] (169,180.19) .. controls (170.43,181.27) and (171.33,182.89) .. (171.34,184.7) .. controls (171.34,185.45) and (171.19,186.16) .. (170.91,186.82) -- (164.79,184.72) -- cycle ; \draw   (169,180.19) .. controls (170.43,181.27) and (171.33,182.89) .. (171.34,184.7) .. controls (171.34,185.45) and (171.19,186.16) .. (170.91,186.82) ;  
        \draw  [dash pattern={on 0.84pt off 2.51pt}]  (153,83.15) -- (178,82.87) ;
        \draw [shift={(178,82.87)}, rotate = 179.35] [color={rgb, 255:red, 0; green, 0; blue, 0 }  ][line width=0.75]    (0,5.59) -- (0,-5.59)   ;
        \draw [shift={(153,83.15)}, rotate = 179.35] [color={rgb, 255:red, 0; green, 0; blue, 0 }  ][line width=0.75]    (0,5.59) -- (0,-5.59)   ;
        \draw  [dash pattern={on 0.84pt off 2.51pt}]  (184,122.87) -- (248,106.87) ;
        \draw [shift={(248,106.87)}, rotate = 165.96] [color={rgb, 255:red, 0; green, 0; blue, 0 }  ][line width=0.75]    (0,5.59) -- (0,-5.59)   ;
        \draw [shift={(184,122.87)}, rotate = 165.96] [color={rgb, 255:red, 0; green, 0; blue, 0 }  ][line width=0.75]    (0,5.59) -- (0,-5.59)   ;
        
        \draw (232.77,150.88) node  [font=\large,color={rgb, 255:red, 255; green, 255; blue, 255 }  ,opacity=1 ] [align=left] {\begin{minipage}[lt]{17.44pt}\setlength\topsep{0pt}
        C
        \end{minipage}};
        \draw (439.26,189.34) node  [font=\large,color={rgb, 255:red, 255; green, 255; blue, 255 }  ,opacity=1 ] [align=left] {\begin{minipage}[lt]{26.64pt}\setlength\topsep{0pt}
        U
        \end{minipage}};
        \draw (83.41,83.15) node  [font=\large,color={rgb, 255:red, 255; green, 255; blue, 255 }  ,opacity=1 ] [align=left] {\begin{minipage}[lt]{38.88pt}\setlength\topsep{0pt}
        $\displaystyle i^{*} +1$
        \end{minipage}};
        \draw (455.91,75.15) node  [font=\large,color={rgb, 255:red, 255; green, 255; blue, 255 }  ,opacity=1 ] [align=left] {\begin{minipage}[lt]{17.8pt}\setlength\topsep{0pt}
        $\displaystyle i^{*}$
        \end{minipage}};
        \draw (309.2,126.9) node [anchor=north west][inner sep=0.75pt]  [font=\large]  {$\phi $};
        \draw (158,158.4) node [anchor=north west][inner sep=0.75pt]  [font=\large]  {$\theta $};
        \draw (320.2,152.4) node [anchor=north west][inner sep=0.75pt]  [font=\large,color={rgb, 255:red, 208; green, 2; blue, 27 }  ,opacity=1 ]  {$v( t)$};
        \draw (160,60.4) node [anchor=north west][inner sep=0.75pt]    {$\epsilon _{v}$};
        \draw (198,91.55) node [anchor=north west][inner sep=0.75pt]    {$L_{w}$};

        \end{tikzpicture}
    \end{adjustbox}
    \caption{Vehicle Dynamics Diagram}
    \label{fig:lateral_model}
     \vspace*{-1.5\baselineskip}
\end{figure}

\textbf{Vehicle Dynamics:}
Every CAV $i\in \mathcal{S}(t)$ follows the dynamics of a control-affine approximated kinematic bicycle model, as defined in \cite{he2021rule}:
\begin{equation}
\label{eq:vehicle_full_dynamics}
    \resizebox{\linewidth}{!}{$
    \underbrace{\left[\begin{array}{c}
    \dot{x} \\
    \dot{y} \\
    \dot{\theta} \\
    \dot{v}
    \end{array}\right]}_{\mathbf{\dot{x}}}
    =
    \underbrace{\left[\begin{array}{c}
    v \cos (\theta) \\
    v \sin (\theta) \\
    0 \\
    0
    \end{array}\right]}_{f\left(\mathbf{x}(t)\right)}+
    \underbrace{\left[\begin{array}{cc}
    0 & -v \sin (\theta) \\
    0 & v \cos (\theta) \\
    0 & v / L_w \\
    1 & 0
    \end{array}\right]}_{g\left(\mathbf{x}(t)\right)}
    \underbrace{\left[\begin{array}{l}
    u \\
    \phi
    \end{array}\right]}_{\mathbf{u}(t)}
    +\underbrace{H{\omega}(t)}_{d(t,\mathbf{x})}$}
\end{equation}
 where the state variables $x$, $y$, $\theta$, and $v$ represent the longitudinal position, lateral position, heading angle, and speed, respectively. Similarly, the control inputs are vehicle acceleration $u$ and steering angle $\phi$. The disturbance $d(t,x)$ is an additive bounded noise represented by the vector ${\omega}(t)\in W\subset \mathbb{R}^q$, where $W$ is a defined subset of $\mathbb{R}^q$, and $H\in\mathbb{R}^{4\times q}$ is a mapping matrix of appropriate dimensions. The physical interpretation of the variables in \eqref{eq:vehicle_full_dynamics} is illustrated in Fig. \ref{fig:lateral_model}.

Note that on a straight segment with $\theta=0$, $\phi=0$, and no disturbances, the system reduces to double integrator dynamics of the form 
\begin{equation}
    \dot{x}_i(t) = v_i(t), \;
    \dot{v}_i(t) = u_i(t)        
    \label{eq:vehicle_long_dynamics}
\end{equation}

 Thus, for purposes of generating a reference longitudinal trajectory for the maneuver, we assume that every vehicle travels in a straight line under the dynamics described in \eqref{eq:vehicle_long_dynamics}. The maneuver is initiated at time $t_{0}$ and the completion time of the longitudinal maneuver is denoted as $t_f$. The control and speed variables have the following constraints:
\begin{subequations}
    \begin{align}
    u_{i_{\min}}\leq u_i(t)\leq u_{i_{\max}}, &\quad \forall t\in\lbrack t_{0},t_{f}\rbrack \label{subeq:cntrol_cnstraints}\\
    v_{i_{\min}}\leq v_i(t)\leq v_{i_{\max}}, &\quad \forall t\in\lbrack t_{0},t_{f}\rbrack \label{subeq:speed_cnstraints}
    \end{align}
    \label{eq:vehicle_constraints}
\end{subequations}
In the above expressions, $v_{i_{\max}}>0$ and $v_{i_{\min}}>0$ represent the maximum and minimum speeds allowed on the highway, while $u_{i_{\max}}>0$ and $u_{i_{\min}}<0$ represent the maximum and minimum acceleration controls for CAV $i$. 

\textbf{Rear-End Safety Constraints:} 
Given a vehicle $i$ and its immediately preceding vehicle $i_p$, we define the minimum safety distance of $i$ with respect to $i_p$ as a speed-dependent distance constraint of the form
\begin{equation}
x_{i_p}(t)-x_i(t) \geq \varphi v_i(t)+\varepsilon \quad \forall t\in\lbrack t_{0},t_{f}\rbrack
\label{eq:safety_constraint}
\end{equation}
where $\varepsilon$ represents a fixed constant offset and $\varphi$ denotes the reaction time (as a rule, $\varphi = 1.8$ is suggested, see \cite{vogel2003comparison}).

\textbf{Traffic Disruption:}
We aim to quantify the impact of a lane-changing maneuver on fast-lane traffic. Let $t_f>t_0$ and $x_i(t_f)$ be the final position of CAV $i$ under control policy $u_i(t)$ for $t\in \lbrack t_0,t_f]$. At time $t>t_0$, we define a disruption metric $D_i(t)$ for any CAV $i \in S(t)$ as follows:
\begin{subequations}
\label{eq:disruption-metric-introduction}
    \begin{gather}
        \mathcal{D}_i(t)=\gamma_x \mathcal{D}_i^x(t)+\gamma_v \mathcal{D}_i^v(t), \\
        \mathcal{D}_i^x(t)=\left(x_i(t)-\left(x_i(0)+v_i(0) \cdot t\right)\right)^2, \\
        \mathcal{D}_i^v(t)=\left(v_i(t)-v_d\right)^2,\\
        \gamma_x =\gamma\cdot\left(\max \left(v_{\max }-v_0, v_{\min }-v_0\right) \cdot t_{a v g}\right)^{-2} \\
        \gamma_v =(1-\gamma) \cdot \max \left(v_{\max }-v_d, v_{\min }-v_d\right)^{-2}
    \end{gather}
\end{subequations}
Here, $\mathcal{D}_i^{x}(t)$ and $\mathcal{D}_i^{v}(t)$ represent disruption metrics for position and speed, respectively. The position disruption $\mathcal{D}_i^x(t)$ quantifies the deviation in the final positions of vehicles in the fast lane compared to where they would have been if they had maintained their initial speed over $[0, t]$ without cooperating. The flow speed disruption $\mathcal{D}_i^v(t)$ measures the deviation of a vehicle's speed at time $t$ from a desired flow speed, $v_d$. The term $t_{avg}$ represents the average longitudinal maneuver time. The weights $\gamma_x$ and $\gamma_v$ in \eqref{eq:disruption-metric-introduction} normalize and combine each disruption terms in a convex manner. The parameter $\gamma\in [0,1]$ adjusts the relative importance of each disruption type.  For further details, see \cite{armijos2022cooperative}.

\textbf{Minimally Disruptive Terminal Conditions:}
We summarize the findings from our previous work \cite{chalaki2023minimally} where we seek to find the optimal longitudinal maneuver time $t_f^*$, together with the longitudinal terminal position $x_i^*(t_f^*)$ for each CAV $i \in \mathcal{S}(t_f^*)$, where $\mathcal{S}(t_f^*)$ denotes the relevant CAV set with respect to CAV $C$ at time $t_f^*$. Specifically, we seek to create a  minimally disruptive safe gap between a CAV pair $i^*$ and $i^*+1$ such that CAV $C$ can perform a lateral maneuver. Thus, given a maximum allowable terminal terminal $T_{\max}$, the terminal conditions for the longitudinal maneuver segment can be determined by solving a mixed integer nonlinear programming problem (MINLP) of the form
 \begin{subequations}
    \label{eq:terminal_states}
     \begin{equation}
        \argmin_{t_f,B_i,\textbf{x}_i(t_f),{v}_i(t_f)} \left\{ \frac{1}{T_{\max}} t_f + \sum_{i\in \mathcal{S}(t_f)\cup\{C\}}D_i(t_f) \right\} \nonumber\\ 
     \end{equation}
     \vspace*{-0.5\baselineskip}
    \begin{gather}
        \textit{s.t.}\;  x_U\left(t_f\right)-x_C\left(t_f\right) \geq \varphi v_C\left(t_f\right)+\epsilon \label{sub_eq:safety_CU}\\
         \forall j \in \mathcal{S}(t) \backslash\{N\}: \notag \\
         x_j\left(t_f\right)-x_{j+1}\left(t_f\right) \geq \varphi v_{j+1}\left(t_f\right)+\epsilon \label{sub_eq:safety_j_preceding}\\
         \forall i \in \mathcal{S}(t): \nonumber \\
         x_C\left(t_f\right)-x_i\left(t_f\right)+\left(1-B_i\right) M \geq \varphi v_i\left(t_f\right)+\epsilon \label{sub_eq:safety_binary_cj_true} \\
         x_i\left(t_f\right)-x_C\left(t_f\right)+B_i M \geq \varphi v_C\left(t_f\right)+\epsilon, \label{sub_eq:safety_binary_cj_false}\\
         \forall i \in \mathcal{S}(t) \cup\{C\}: \nonumber \\
         p^{\mathrm{u}}\left(x_i, v_i, t_f\right) \leq 0 \cup p^{\text {l }}\left(x_i, v_i, t_f\right) \geq 0 \\
         0 \leq v_{\min } \leq v_i\left(t_f\right) \leq v_{\max },
    \end{gather}
\end{subequations}
where \eqref{sub_eq:safety_CU}-\eqref{sub_eq:safety_binary_cj_false} describe the rear-end safety constraints as defined in \eqref{eq:safety_constraint}. Additionally, $B_i$ is a binary variable that determines if the optimal CAV $C$ should merge ahead of CAV $i$. Furthermore, $p^{\mathrm{u}}$ and $p^{\mathrm{l}}$ represent the reachable set for a point mass model \cite{haddad2022boundary}, and they are defined by the following equations:
\begin{subequations}
    \begin{gather}
        \resizebox{0.85\linewidth}{!}{$p^{\mathrm{u}}\left(x_i, v_i, t\right)=  -\frac{t^2}{2}+\frac{1}{4}\left(\frac{v_i-v_i(0)-\nu t}{\mu}+t\right)^2 - \mathcal{R}$}\\
        \resizebox{0.85\linewidth}{!}{$p^{\mathrm{l}}\left(x_i, v_i, t\right)=  \frac{t^2}{2}-\frac{1}{4}\left(\frac{-v_i+v_i(0)+\nu t}{\mu}+t\right)^2 - \mathcal{R}$}\\
        \mathcal{R} = \mu^{-1}\left(x_i-x_i(0)-t v_i(0)-0.5\nu {t^2}\right)\\
        \mu=0.5\left({u_{\max }-u_{\min }}\right),\;\nu=0.5\left(u_{\max }+u_{\min }\right) 
    \end{gather}
    \label{eq:reachable_set}
\end{subequations} 
\textbf{Decentralized Optimal Longitudinal Trajectory:}
The longitudinal trajectory is defined once the terminal time $t_f^*$, terminal position $x^*_{i,f}$, and terminal speed $v^*_{i,f}$ for each CAV $i$ are computed by solving \eqref{eq:terminal_states}. These are then used in defining the following OCP for determining $u_i(t)$:
\begin{subequations}
\label{eq:ocp_cav}
\begin{gather}
\min \limits_{u_{\min}\leq u_i(t) \leq u_{\max}} \int_{t_0}^{t_f^*} \frac{1}{2} u_i^2(t) d t \label{sub_eq:ocp_cav_obj}\\
\text { s.t. \eqref{eq:vehicle_long_dynamics}, \eqref{eq:vehicle_constraints}}, \label{sub_eq:ocp_cav_dyn}\notag\\
x_{i-1}(t)-x_i(t) \geq \varphi v_i(t)+\varepsilon, \quad \forall t \in\left[t_0, t_f^*\right] \label{sub_eq:ocp_cav_safety_cons} \\
x_i\left(t_f^*\right)=x^*_{i,f}, \quad
v_i\left(t_f^*\right)=v^*_{i,f} \label{sub_eq:ocp_terminal_vel_cons}
\end{gather}
\end{subequations} 

\textbf{Lateral Maneuver:}
The lateral maneuver is initiated at time $t_0^l$ when CAV $C$ is longitudinally safe with respect to vehicle $U$ and the two vehicles ahead and behind the optimal merging pair $(i^*, i^{*}+1)$ obtained from solving \eqref{eq:terminal_states}. The value of $t_0^l$ is computed based on a minimum safe distance $\varepsilon_l$. Specifically, $t_0^l$ is determined as follows:
\begin{equation}
    \begin{gathered}
        t_0^l = \min \left\{t\in[t_0,t_f^*]:\left(x_{U}(t)-x_{C}(t)\geq \varepsilon_l \right) \land\right.   \\
                \left. \left(x_{i^*}(t)-x_{C}(t)\geq \varepsilon_l\right) 
                 \land \left(x_{C}(t)- x_{i^*+1}(t)\geq \varepsilon_l) \right) \right\}.
    \end{gathered}
    \label{eq:lane_changing_conditions}
\end{equation}
Similarly, we define the end of the lateral maneuver $t_f^l$ as the time at which CAV $C$ reaches the center of the fast lane, denoted as $y_{des}$. It is important to note that under ideal conditions, $t_0^l\leq t_f^*$ since $\varepsilon_l$ can be smaller than the minimum longitudinal safety specified in \eqref{eq:safety_constraint}.


 \textbf{Problem Formulation with temporal constraints and bounded disturbances:}
Note that the solution to \eqref{eq:ocp_cav} assumes perfect knowledge of the preceding vehicle's state, making the problem challenging when the behavior of this vehicle is unknown or time-varying. Additionally, the presence of external disturbances can render the solutions infeasible. Furthermore, the nonlinearity of the dynamics prevents the guarantee of global optimality for the lateral control trajectory. Therefore, in this paper, we address these challenges and formulate the following problem:

\begin{prob}
Given system \eqref{eq:vehicle_long_dynamics} subject to bounded disturbances $d:|d(t,x)|\leq \gamma$ and potential variations in the behavior of uncooperative vehicles $U$ and $F$, our objective is to derive an online optimal control policy $u^*(t)$ that ensures the safety of CAV $i$ with respect to others, as defined in \eqref{eq:safety_constraint}, while deviating minimally from the solutions of the original OCPs defined in \eqref{eq:ocp_cav}.
\end{prob}

%% file: sections/planning2execution.tex
\textbf{Unconstrained optimal control solution:}
Based on problem \eqref{eq:ocp_cav}, we derive the analytical optimal control solution such that any CAV $i \in \mathcal{S}(t)\cup \{C\}$ reaches the specified terminal conditions defined by \eqref{eq:terminal_states} and described in constraint \eqref{sub_eq:ocp_terminal_vel_cons}. Let $\mathbf{x}_i(t):=(x_i(t),v_i(t))^T$ and $\mathbf{\lambda}_i(t)=(\lambda_i^x(t),\lambda_i^v(t))^T$ be the state and costate variables, respectively. We define    the Hamiltonian of \eqref{eq:ocp_cav} as:
\begin{multline}
    \label{eq:hamiltonian} H(\mathbf{x}_i,\mathbf{\lambda}_i,u_i)=
    \frac{1}{2}u_i^2+\lambda_i^x v_i+\lambda_i^v u_i +\mu_1(v_{i_{\min}}-v_i)\\
    +\mu_2(v_i-v_{i_{\max}})+\mu_3(u_{i_{\min}}-u_i)+
    \mu_4(u_i-u_{i_{\max}})\\
    +\mu_5 (x_i(t)+\varphi v_i(t)+\epsilon-x_{i-1}(t)).
\end{multline}

The Lagrange multipliers $\mu_1,\mu_2,\mu_3,\mu_4,\mu_5$ are positive when their corresponding constraints are active and become 0 when the constraints are inactive. The terminal position and terminal speed of CAV $i$ are specified at the terminal time $t_f^*$ by functions $\psi_1:=x_i(t_f^*)-x_i^f=0$ and $\psi_2:=v_i(t_f^*)-v_i^f=0$, respectively. Furthermore, the Euler-Lagrange equations provide:
\begin{equation}
\label{eq:eular_lagrange}
\dot{\lambda}_i^x=-\dfrac{\partial H}{\partial x_i}= -\mu_5,\; \;
\dot{\lambda}_i^v=-\dfrac{\partial H}{\partial v_i}=-\lambda_i^x+\mu_3-\mu_4-\varphi\mu_5,
\end{equation}
and the necessary conditions for optimality are
\begin{align}
\label{eq:optimality_condition}
\dfrac{\partial H}{\partial u_i}=u_i(t)+\lambda_i^v(t)-\mu_1+\mu_2=0.
\end{align}

In the unconstrained case, the Lagrange multipliers satisfy $\mu_1=\mu_2=\mu_3=\mu_4=\mu_5=0$. Therefore, we have $\dot{\lambda}_i^x=0$ and $\dot{\lambda}_i^v=-\lambda_i^x$ from \eqref{eq:eular_lagrange}, and we can further obtain $\lambda_i^x=a_i$ and $\lambda_i^v=-(a_it+b_i)$, where $a_i$ and $b_i$ are integration constants. The optimality condition \eqref{eq:optimality_condition} provides $u_i(t)=-\lambda_i^v$, which leads to the following optimal solution:

\begin{subequations}
\label{eq:unconstrained_sol_system}
    \begin{align}
        &u^*_i(t)=a_it+b_i,\\
        &v_i^*(t)=\frac{1}{2}a_it^2+b_it+c_i,\\
        &x_i^*(t)=\frac{1}{6}a_it^3+\frac{1}{2}b_it^2+c_it+d_i,
    \end{align}
\end{subequations}
where $c_i$ and $d_i$ are integration constants. Combining the optimal solution in \eqref{eq:unconstrained_sol_system} with the known initial and terminal states, we can solve for the constants $a_i$, $b_i$, $c_i$, and $d_i$ by substituting $t=t_0$ and $t=t^*_f$ into \eqref{eq:unconstrained_sol_system} and equating the results to the corresponding initial and terminal states:
\begin{subequations}
    \label{eq:unconstrained_sol}
    \begin{align}
    &v_i^*(t_0)=\frac{1}{2}a_it_0^2+b_it_0+c_i,\\
    &x_i^*(t_0)=\frac{1}{6}a_it_0^3+\frac{1}{2}b_it_0^2+c_it_0+d_i,\\
    &v_i^*(t_f)=\frac{1}{2}a_it_f^2+b_it_f+c_i,\\
    &x_i^*(t_f)=\frac{1}{6}a_it_f^3+\frac{1}{2}b_it_f^2+c_it_f+d_i,
\end{align}
\end{subequations}
By solving this system of equations, we can determine the values of $a_i$, $b_i$, $c_i$, and $d_i$, allowing us to obtain the analytical expressions for the unconstrained optimal solution $x_i^*(t)$, $v_i^*(t)$, and $u_i^*(t)$ given by \eqref{eq:unconstrained_sol_system}.



    
\textbf{Reference Control Input:}
In order to optimally drive each CAV $i$ in $\mathcal{S}(t)\cup\{C\}$ from its initial conditions to its optimal terminal position $x_{i,f}^*$, we compute a control solution that tracks the solution obtained from the OCP in \eqref{eq:unconstrained_sol}. This is accomplished by defining a feedback law using a mapping function $\Omega:\mathbb{R}\rightarrow\mathbb{R}$, which maps the current observed position $\hat{x}_i(t)$ at time $t$ to the nearest reference position in the OCP \eqref{eq:unconstrained_sol}. This mapping allows us to determine the control reference $u_{\text{ref}}$ and the reference speed $v_{\text{ref}}$ using the functions $\Omega_v(x(t))$ and $\Omega_u(x(t))$, respectively. Thus, we have
\begin{equation}
        v^i_{\text{ref}} = \Omega_v(\hat{x}_i(t)), \quad
        u^i_{\text{ref}} = \Omega_u(\hat{x}_i(t))
    \label{eq:mapping_reference_function}
\end{equation}
\textbf{Spatio-temporal Constraints:}
 We seek to reach the terminal position $\boldsymbol{x}^*_{i,f}$ by the optimal terminal time $t_f^*$, as computed by \eqref{eq:terminal_states}. In other words, we require that $\lim \limits_{t\rightarrow t_f^*}{x}_i(t)=x^*_{i,f}$ while maintaining the safety constraint \eqref{sub_eq:ocp_cav_safety_cons}. To achieve this objective by $t_f^*$, we use Theorem \ref{thm:conditions_safety_convergence} to define the following goal set with fixed-time convergence:
\begin{equation}
    \mathcal{V}^i_{x_f}(x_i(t)) = \left(x_i(t) - x_{i,f}^* \right)^2
    \label{eq:position_clf_1st_deg}
\end{equation}

It can be observed that the control input $u(t)$ does not appear in the Lie derivative of \eqref{eq:position_clf_1st_deg} along the system dynamics \eqref{eq:vehicle_full_dynamics}. This is commonly encountered and can be addressed (see \cite{xiao2023safe}) by redefining $\mathcal{V}^i_{x_f}(x_i(t))$ and including its first order derivative as
\begin{equation}
    \begin{aligned}
       \mathcal{V}^i_{x_f}(x_i(t)) = &p_1\left(x_i(t) - x_{i,f}^* \right)^2 + \\
                            &2v_i(t)\cos(\theta(t))(x_i(t) - x_{i,f}^*),  
    \end{aligned}
    \label{eq:position_clf}
\end{equation}
where $p_1>0$ is a linear class-$\mathcal{K}$ function gain. The FxT-CLF-$S_G$ constraint can be expressed as: 
\begin{align}
 \mathcal{L}_g  \mathcal{V}^i_{x_f}&(x_i(t))\boldsymbol{u} + \mathcal{L}_f  \mathcal{V}^i_{x_f}(x_i(t)) +
   \sup \limits_{\omega\in W}\left\{\mathcal{L}_d \mathcal{V}^i_{x_f}(x_i(t))\omega \right\}\notag \\
  & \leq -\boldsymbol{\delta}_{x_f}\mathcal{V}^i_{x_f}(x_i(t)) -
   \alpha_1 \max \left\{0,\mathcal{V}^i_{x_f}(x_i(t))\right\}^{\gamma_1}\notag\\ 
  & -\alpha_2 \max \left\{0,\mathcal{V}^i_{x_f}(x_i(t))\right\}^{\gamma_2},\label{eq:fxt_cnstraint_position}
\end{align}
where $\alpha_1$, $\alpha_2$, $\gamma_1$, and $\gamma_2$ are fixed parameters as defined in \eqref{eq:conditions_safety_convergence}. Specifically, we set $\alpha_1=\alpha_2=\frac{\mu \pi}{2 T_{ud}}$, $\gamma_1=1+\frac{1}{\mu}$, and $\gamma_2=1-\frac{1}{\mu}$, where $\mu>1$. In this case, $T_{ud} = \max\left\{\mathcal{T}_{\min},t_f^*-t\right\}$ with $\mathcal{T}_{\min}$ being a small value such that singularities can be avoided. In \eqref{eq:fxt_cnstraint_position}, $\delta_{x_f}$ is an optimization variable that adjusts based on the problem's feasibility at each time step. Furthermore, to satisfy \eqref{eq:fxt_cnstraint_position} under worst-case disturbance, we include the supremum of the term $\mathcal{L}_d \mathcal{V}^i_{x_f}=\frac{\partial \mathcal{V}^i_{x_f}}{\partial x}d(t,x)$, which denotes the Lie derivative of $\mathcal{V}^i_{x_f}$ along the disturbance terms.

\textbf{Soft Constraints:}
To ensure the optimality of the solution, we aim to track the optimal states $x_i^*(t)$ and $v_i^*(t)$ obtained in \eqref{eq:unconstrained_sol} for each CAV $i \in \mathcal{S}(t)\cup \{C\}$. However, the control goals in this case, will only be satisfied when the hard constraints are satisfied. To achieve this, we use the reference states computed with \eqref{eq:mapping_reference_function} to define a Lyapunov function $\mathcal{V}_{v}(\boldsymbol{x}_i(t))$ that tracks the desired optimal reference speed $v^i_{\text{ref}}(t)$ as follows:
\begin{equation}
    \mathcal{V}_{v_{ref}}(\boldsymbol{x}_i(t)) = \left(v_i(t)-v^i_{\text{ref}}(t)\right)^2 
    \label{eq:clf_speed}
\end{equation}

In addition, we wish to minimize changes in the relative heading $\theta_i(t)$ and to maintain the desired lateral position $y^i_{\text{des}}(t)$. This results in the following Lyapunov functions:
\begin{gather}
    \mathcal{V}_{\theta}(\boldsymbol{x}_i(t)) = \left(\theta_i(t)\right)^2 \label{eq:clf_theta}\\ 
    \mathcal{V}_{y_{des}}(\boldsymbol{x}_i(t)) = \left(y_i(t)-y^i_{des}\right)^2,\label{eq:clf_y_des}
\end{gather}
where the desired lateral position $y^i_{des}$ for all of the maneuver time $t\in \left[t_0,t_f \right]$, except for CAV $C$ during the lateral maneuver portion when $t\geq=t_0^l$. In this case, $y^C_{des}=l$, where $l$ denotes the lane width.

Unlike \eqref{eq:fxt_cnstraint_position}, we do not have a terminal time convergence requirement for constraints \eqref{eq:clf_speed}-\eqref{eq:clf_y_des}. Instead, we allow the relaxation of the corresponding CLF constraints by including a slack variable $e_i$ for \eqref{eq:clf_speed}-\eqref{eq:clf_y_des} as follows:
\begin{equation}
    \begin{gathered}
        \forall j \in \left\{ v_{ref}, \theta, y_{des}\right\}\\
        \mathcal{L}_g  \mathcal{V}_j(x_i(t))\boldsymbol{u} + \mathcal{L}_f  \mathcal{V}_j(x_i(t))+\epsilon_j\mathcal{V}_j(x_i(t)) +\\
        \sup\limits_{\omega\in W}\left\{\mathcal{L}_d \mathcal{V}_j(x_i(t))\omega \right\} \leq \boldsymbol{e}_j, 
    \end{gathered}
    \label{eq:clf_constraint_objectives}
\end{equation}

\textbf{Hard Constraints:}
To account for safety constraints in cooperative lane-changing maneuvers, we introduce two types of safety constraints. The first is the safe distance with respect to the immediately preceding vehicle. Specifically, we define the CBF for \eqref{sub_eq:ocp_cav_safety_cons} as follows:
    \begin{equation}
        \label{eq:cbf_preceding_safety}
        h_{x_p}(\boldsymbol{x}_i)=x_{i-1}(t)-x_{i}(t)-\varphi v_i(t)-\epsilon
    \end{equation}
where $x_{i-1}$ denotes the position of the vehicle immediately ahead of the ego CAV $i$.

The second safety constraint pertains to the safe merging constraint. In some cases, the region of attraction defined by \eqref{eq:position_clf} in \eqref{eq:fxt_cnstraint_position} may not be feasible due to changes in the speed of an uncooperative vehicle $U$ ahead of $i$, which could violate safety constraint \eqref{eq:cbf_preceding_safety}. To address this issue, we define an additional constraint that ensures the feasibility of the maneuvers when fixed-time convergence is not possible. This constraint only affects CAV $C$ and CAV $i_m$, where $i_m$ denotes the first vehicle in $\mathcal{S}(t)$ that must decelerate so that CAV $C$ can merge ahead (e.g., CAV $2$ in Fig. \ref{fig:lateral_model}). Inspired by \cite{xu2021decentralized}, we define the following:

\begin{defn}\label{def:reaction_time}
The reaction time $\varphi$ of CAV $i$ relative to an adjacent vehicle $i_a$ is a smooth and strictly increasing function  $\Phi: \mathbb{R} \rightarrow \mathbb{R}$ with initial condition $\Phi\left(x_i\left(t_0\right)\right)=-\frac{\epsilon}{v_i(t_0)}$ and final condition $\Phi\left(x_i\left(t_m\right)\right)=\varphi$, where $t_0$ denotes the initial maneuver time and $t_m$ denotes the time at which the lateral maneuver takes place for $C$.
\end{defn}

We can now define the safe merging function as
\begin{equation}
    h_{m}(\boldsymbol{x}_i)=x_{i_a}(t)-x_{i}-\Phi\left(x_i(t)\right) v_i(t)-\epsilon \geq 0, \; \forall t \in\left[t_0, t_f^*\right]
    \label{eq:merging_feasibility_const}
\end{equation}
where the index $i_a$ represents an adjacent vehicle traveling on the lane adjacent to $i$ (e.g., $i=2$, $a=C$ in Fig. \ref{fig:lateral_model}). Lastly, let $\Phi: \mathbb{R} \rightarrow \mathbb{R}$ denote a strictly increasing smooth mapping function that guarantees that $\Phi(t_f^*)=\varphi$ such that \eqref{eq:merging_feasibility_const} is feasible at time $t_f^*$ as defined in Def. \ref{def:reaction_time}. Thus, we define the function $\Phi$ as:
\begin{equation}
    \resizebox{0.89\linewidth}{!}{$\Phi\left(x_i(t)\right)=\left(\varphi+\frac{L_{i_a}-L_i+\epsilon}{v_i\left(t_0\right)}\right) \frac{x_i(t)}{L_i}-\frac{L_{i_a}-L_i+\epsilon}{v_i\left(t_0\right)}
    \label{eq:monotonic_merging_function}$}
\end{equation}
where $L_{i_a}$ and $L_i$ denote constants of the form $L_{i_a} = x^{*}_{C_f} - x_a(t_0)$ and $L_{i} = x^{*}_{C_f} - x_i(t_0)$, respectively.

Similarly, we define two additional CBFs to account for constraint \eqref{subeq:speed_cnstraints} as:
\begin{gather}
    h_{v_{\min}}\left(\boldsymbol{x}_i(t)\right) = v_i(t)-v_{\min} \label{eq:v_min_cbf}\\
    h_{v_{\max}}\left(\boldsymbol{x}_i(t)\right) = v_{\max}-v_i(t) \label{eq:v_max_cbf}
\end{gather}

Using \eqref{eq:cbf_preceding_safety}, \eqref{eq:merging_feasibility_const}, \eqref{eq:v_min_cbf}, and \eqref{eq:v_max_cbf} we define the CBF constraints as: 
\begin{equation}
    \begin{gathered}
        \forall j \in  \left\{ x_p, m, v_{\min}, v_{\max}\right\}\\
        \resizebox{0.89\linewidth}{!}{$
        \mathcal{L}_g h_j(x) \boldsymbol{u} + \mathcal{L}_f h_j(x) +\boldsymbol{\delta}_j h_j(x)  \geq  \inf \limits_{\omega\in W}\left\{\mathcal{L}_d h_j(x_i(t))\omega \right\}\ $}
    \end{gathered}   
    \label{eq:cbf_constraints}
\end{equation}
where similar to \eqref{eq:fxt_cnstraint_position} and \eqref{eq:clf_constraint_objectives}, we compute the lie derivative of $h_j$ along the disturbance terms so as to account for the worst-case disturbance possibility.

\textbf{Disturbance Rejection:}
Based on the problem formulation, the robustification of each constraint involves considering the worst-case disturbance. This is done by solving linear programming problems to compute the supremum and infimum of certain terms involving the disturbance. Specifically, the terms $\sup \limits_{\omega\in W}\left\{\mathcal{L}_d \mathcal{V}_j(x_i(t))\omega \right\}$  in \eqref{eq:fxt_cnstraint_position} and \eqref{eq:clf_constraint_objectives}, and $\inf \limits_{\omega\in W}\left\{\mathcal{L}_d h_j(x_i(t))\omega \right\}$ in \eqref{eq:cbf_constraints}, require solving the following linear programming (LP) problems:
\begin{equation}
\label{eq:supremum_infimum_lp_problem}
\begin{aligned}
&\inf _\omega  \mathcal{L}_{d} h_j(x_i(t))\omega, \\
&\text { s.t. }  A \omega \leq b
\end{aligned} \quad
\begin{aligned}
&\sup _\omega  \mathcal{L}_{d} \mathcal{V}_j(x_i(t))\omega \\
&\text { s.t. }  A \omega \leq b
\end{aligned}
\end{equation}
where the constraints $A \omega \leq b$ represent the bounds of the disturbance term $\omega$ as linear constraints in matrix form. Instead of solving each LP separately for every constraint, we can redefine each constraint using the dual formulation of their corresponding LP. This allows us to incorporate the worst-case disturbance directly into each constraints. For example, constraint \eqref{eq:fxt_cnstraint_position} can be redefined as follows:
\begin{subequations}
\label{eq:robust_fxt_clf_terminal_position}
    \begin{gather}
       \mathcal{L}_g  \mathcal{V}^i_{x_f}(x_i(t))\boldsymbol{u} + \mathcal{L}_f  \mathcal{V}^i_{x_f}(x_i(t)) + b^\top \lambda_{x_f}  \leq \\
       \nonumber  -\boldsymbol{\delta}_{x_f}\mathcal{V}^i_{x_f}(x_i(t)) -\alpha_1 \max \left\{0,\mathcal{V}^i_{x_f}(x_i(t))\right\}^{\gamma_1}\\
       \nonumber - \alpha_2 \max \left\{0,\mathcal{V}^i_{x_f}(x_i(t))\right\}^{\gamma_2}, \\
       A^{\top}\lambda_{x_f}=\mathcal{L}_{d}\mathcal{V}^i_{x_f}(x_i(t)), \quad \lambda_{x_f} \geq 0
    \end{gather}
\end{subequations}
with $\lambda_{x_f}$ being a dual optimization variable. Similarly, we redefine the CLF constraints \eqref{eq:clf_constraint_objectives} as:
\vspace*{-0.5\baselineskip}
\begin{subequations}
\label{eq:robust_clf_constraints}
    \begin{gather}
       \nonumber \forall j \in \left\{ v_{ref}, \theta, y_{des}\right\}\\
       \resizebox{0.87\linewidth}{!}{$\mathcal{L}_g  \mathcal{V}_j(x_i(t))\boldsymbol{u} + \mathcal{L}_f  \mathcal{V}_j(x_i(t))+\epsilon_j\mathcal{V}_j(x_i(t)) + b^\top \lambda_{j} \leq \boldsymbol{e}_j,$} \\
        A^{\top}\lambda_{x_f}=\mathcal{L}_{d}\mathcal{V}^i_{j}(x_i(t)), \quad \lambda_{j} \geq 0,
    \end{gather}
\end{subequations}
Lastly, for the CBF constraints \eqref{eq:cbf_constraints}, we obtain the following
\vspace*{-\baselineskip}
\begin{subequations}
\label{eq:robust_cbf_constraints}
    \begin{gather}
        \nonumber \forall j \in  \left\{ x_p, m, v_{\min}, v_{\max}\right\} \\
       \mathcal{L}_g h_j(x) \boldsymbol{u} + \mathcal{L}_f h_j(x) +\boldsymbol{\delta}_j h_j(x)  \geq  b^\top \mu_j \\
        A^{\top}\mu_{j}=\mathcal{L}_{d}h_{j}(x_i(t)), \quad \mu_{j} \leq 0,
    \end{gather}
\end{subequations}
with $\mu_j$ defined as a dual decision variable for each of the CBF constraints.

\textbf{Optimal State Tracking:} Finally, in order to provide a minimal deviation from the control input computed in \eqref{eq:unconstrained_sol} and \eqref{eq:unconstrained_sol_system}, we use \eqref{eq:qp_cbf_clf_fxt} to construct the OCBF controller with spatio-temporal constraints (FxT-OCBF) as the following optimization problem.
\begin{gather}
\resizebox{1\linewidth}{!}{$\min \limits_{z_i(t),\lambda_{x_f}\geq 0,\lambda_{j}\geq 0, \mu_{k}\leq 0} \int_{t_0}^{t_f}\left(\frac{1}{2} z^\top(t) Q z_i(t)+F^\top z_i(t) \right) dt$}\notag\\
\text{s.t. } \eqref{eq:robust_fxt_clf_terminal_position}, \eqref{eq:robust_clf_constraints}, \eqref{eq:robust_cbf_constraints}, \eqref{subeq:cntrol_cnstraints}\notag\\
\forall j \in \left\{ v_{ref}, \theta, y_{des}\right\},\quad
\forall k \in \left\{ x_p, m, v_{\min}, v_{\max}\right\} 
\label{eq:objective_qp_lane_change}
\end{gather}
where the decision variable vector $z(t)$ contains the optimization variables of the form $z(t) =$ \resizebox{\linewidth}{!}{$ \left[u_i(t)-u^i_{ref}, \phi_i(t), \delta_{x_f}, \delta_{x_p}, \delta_{m}, \delta_{v_{\min}}, \delta_{v_{\max}}, e_{v_{ref}},e_{\theta}, e_{y_{d}}\right]^\top,$}  where $u^i_{ref}$ is the reference acceleration of CAV $i$, and $u_i(t)$ is the actual acceleration input. The weighting matrices $Q\geq0$ and $F\geq0$ are appropriately sized.

%% file: sections/simulation_results.tex
This section presents simulation results demonstrating the effectiveness of the derived OCBF controllers with spatio-temporal constraints (FxT-OCBF). The proposed controller is compared to two other variations of OCBFs: one without spatio-temporal constraints (OCBF) and one without state feedback, resulting in a simple CBF. Section that we are unaware of any method that provides safety GUARANTEES, including the temporal ones introduced in this paper, as well as disturbance rejection. We could compare to any standard car-following model, as we did in other papers, but we have already shown that our approach us superior to these (ref. [5],[6]). Simulations were conducted using MATLAB and the CasADi optimization modeling framework \cite{Andersson2019}. BONMIN was employed to solve the MINLP problem in \eqref{eq:terminal_states}, while OSQP was used to solve the QP problem in \eqref{eq:objective_qp_lane_change}.

To evaluate each controller, two distinct scenarios were considered to demonstrate their ability to minimize disruption and their robustness against disturbances and changes in the behavior of uncontrolled vehicles. In both scenarios, a set $S(t)$ of four cooperative vehicles is traveling in the fast lane, following an uncooperative vehicle $F$ moving at constant speed. Similarly, a CAV $C$ is traveling in the slow lane, following an uncooperative vehicle $U$ also moving at a constant speed. In the second scenario, an identical case to scenario 1 was introduced, but this time vehicles $U$ and $F$ were allowed to adjust their speed profiles throughout the maneuver. Specifically, a constant deceleration of $2.5\, m/s^2$ was applied to both vehicles until they reached the minimum speed. The initial conditions for each vehicle in each scenario are provided in Table \ref{tab:initial_conditions}.

\begin{table}[hptb]
    \centering
    \caption{Initial Conditions}
    \resizebox{\linewidth}{!}{
    \begin{tabular}{|c|c|c|c|c|c|c|c|}
    \hline
    {\ul \textbf{Description}} &
      {\ul \textbf{F}} &
      {\ul \textbf{U}} &
      {\ul \textbf{1}} &
      {\ul \textbf{2}} &
      {\ul \textbf{3}} &
      {\ul \textbf{4}} &
      {\ul \textbf{C}} \\ \hline
    Long. Position $x(t_0)\, [m]$ & 115 & 100  & 85  & 60  & 25  & 0   & 30   \\ \hline
    Lat. Position $y(t_0)\, [m]$ & 1.8 & -1.8 & 1.8 & 1.8 & 1.8 & 1.8 & -1.8 \\ \hline
    Speed $v(t_0)\, [m/s]$ &
      \multicolumn{1}{l|}{28} &
      \multicolumn{1}{l|}{20} &
      \multicolumn{1}{l|}{28} &
      \multicolumn{1}{l|}{24} &
      \multicolumn{1}{l|}{24} &
      \multicolumn{1}{l|}{24} &
      \multicolumn{1}{l|}{24} \\ \hline
    Heading  $\theta(t_0)\, [deg]$ &
      \multicolumn{1}{l|}{0} &
      \multicolumn{1}{l|}{0} &
      \multicolumn{1}{l|}{0} &
      \multicolumn{1}{l|}{0} &
      \multicolumn{1}{l|}{0} &
      \multicolumn{1}{l|}{0} &
      \multicolumn{1}{l|}{0} \\ \hline
    \end{tabular}}
    \label{tab:initial_conditions}
\end{table}
For both scenarios, we introduced an additive uniformly distributed noise process, denoted as $\omega_x$ and $\omega_v$, with zero mean and standard deviations of $\sigma_x=1\,m$ and $\sigma_v=0.5\,m/s$, respectively. These noise processes affect the position and speed states of the vehicles. The speed limits for each vehicle were defined as $v_{\min} = 10\, m/s$ and $v_{\max} = 35\, m/s$. The control limits for each CAV were specified as $u_{\min} = -7\, m/s^2$ and $u_{\max} = 3.3\, m/s^2$. The minimum and maximum steering angle limits were defined as $\phi_{\min} = -15\, deg$ and $\phi_{\max} = 15\, deg$. The wheelbase length for every vehicle was defined as $L_w = 4 \,m$. The minimum safety distance was set to $\varepsilon = 7\, m$, and the reaction time was defined as $\varphi = 0.8\, s$. Similarly, the minimum safety distance to perform a lane-changing maneuver was $\varepsilon_l=15\,m$. The lane width was defined as $l=3.6\,m$.

To compute the terminal states in problem \eqref{eq:terminal_states}, the disruption weighting factor was set as $\gamma = 0.5$, the average longitudinal maneuver length as $t_{avg} = 8\, s$, and the maximum maneuver time as $T_{\max} = 16\, s$. The desired traffic flow speed was defined as $v_{\text{des}}=28\, m/s$. For the spatio-temporal constraint \eqref{eq:robust_fxt_clf_terminal_position}, we defined the constants $\mu=1.5$, $p_1=3$, and $\mathcal{T}_{\min}=0.1$.

\begin{figure*}[ptb]
    \centering   
    \vspace*{0.5\baselineskip}
    \begin{subfigure}{0.49\linewidth}
    \centering 
      \includegraphics[width=\linewidth, height=5cm]{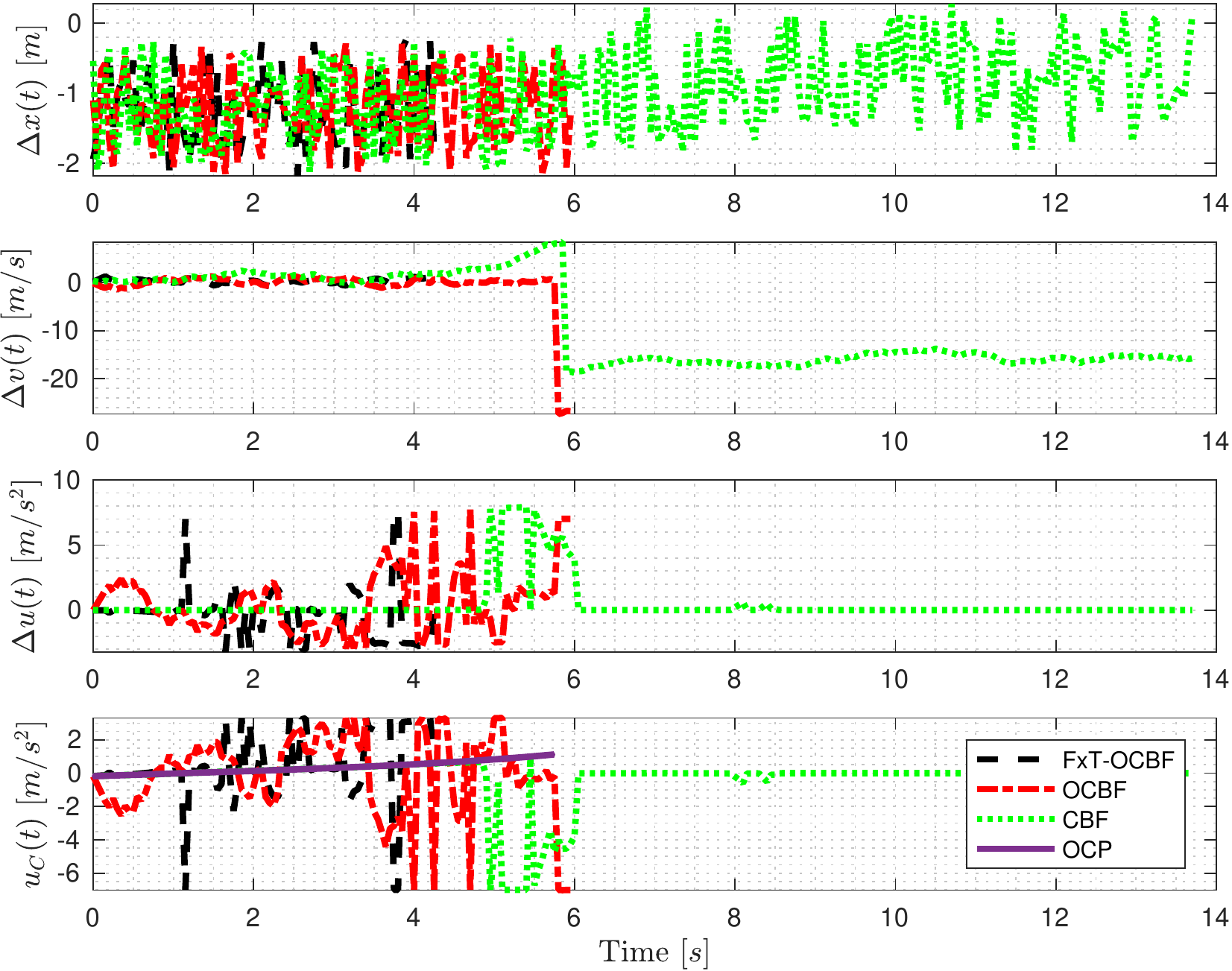}  
      \caption{\centering{State difference divergence for CAV $C$ in scenario 1}}
      \label{sub_fig:diff_ocp_1}
    \end{subfigure}   
    \begin{subfigure}{0.49\linewidth}
    \centering 
      \includegraphics[width=\linewidth, height=5cm]{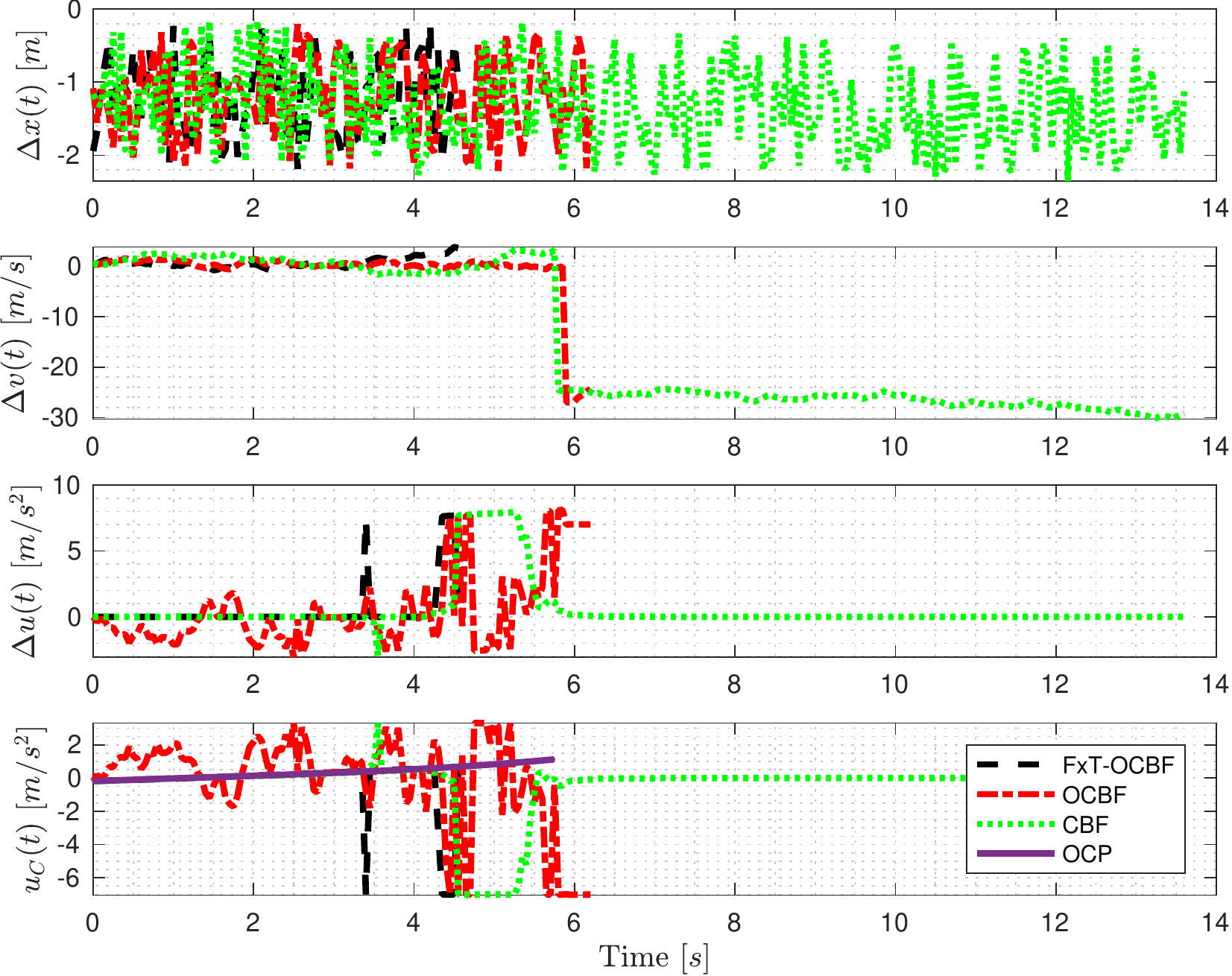}  
      \caption{\centering{State difference divergence for CAV $C$ in scenario 2}}
      \label{sub_fig:diff_ocp_2}
    \end{subfigure}
    \caption{\centering Reference input difference with respect to actual control input and states}
    \label{fig:ocp_difference}
    \vspace*{-\baselineskip}
\end{figure*}

In Fig. \ref{fig:ocp_difference}, we provide a comparison between the different controllers and the analytical unconstrained optimal control solution for the trajectory of CAV $C$. Specifically, we show the difference between the actual trajectory and the reference trajectory computed from solving \eqref{eq:unconstrained_sol} given the initial conditions provided in Table \ref{tab:initial_conditions}. Additionally, we describe the resulting average disruption performance for all CAVs involved in the maneuvers in Table \ref{tab:disruption_results}, and the energy consumption in Table \ref{tab:energy_comparison}.

We will begin by analyzing the first scenario, where the uncooperative vehicles do not experience deceleration during the maneuver. In Fig. \ref{sub_fig:diff_ocp_1}, it can be observed that the signals have different durations due to variations in the total maneuver execution time. The FxT-OCBF method exhibits the shortest duration, which demonstrates the effectiveness of our approach in achieving convergence before $t_f^*$, despite disturbances. The value of $t_f^*$ was computed to be 5.77 s. All three methods closely track the reference speed input. However, for the OCBF and CBF methods, there is a high variation in the speed difference after approximately 5.7 seconds, which coincides with the optimal longitudinal terminal time. This indicates the activation of the rear-end safety constraints, leading to high decelerations in both methods.

\begin{figure}[tpb]
    \vspace*{0.5\baselineskip}
    \centering 
    \label{fig:Trajectory_Sample}
    \begin{subfigure}{1\linewidth}
    \centering 
      \includegraphics[width=\linewidth, height=5cm]{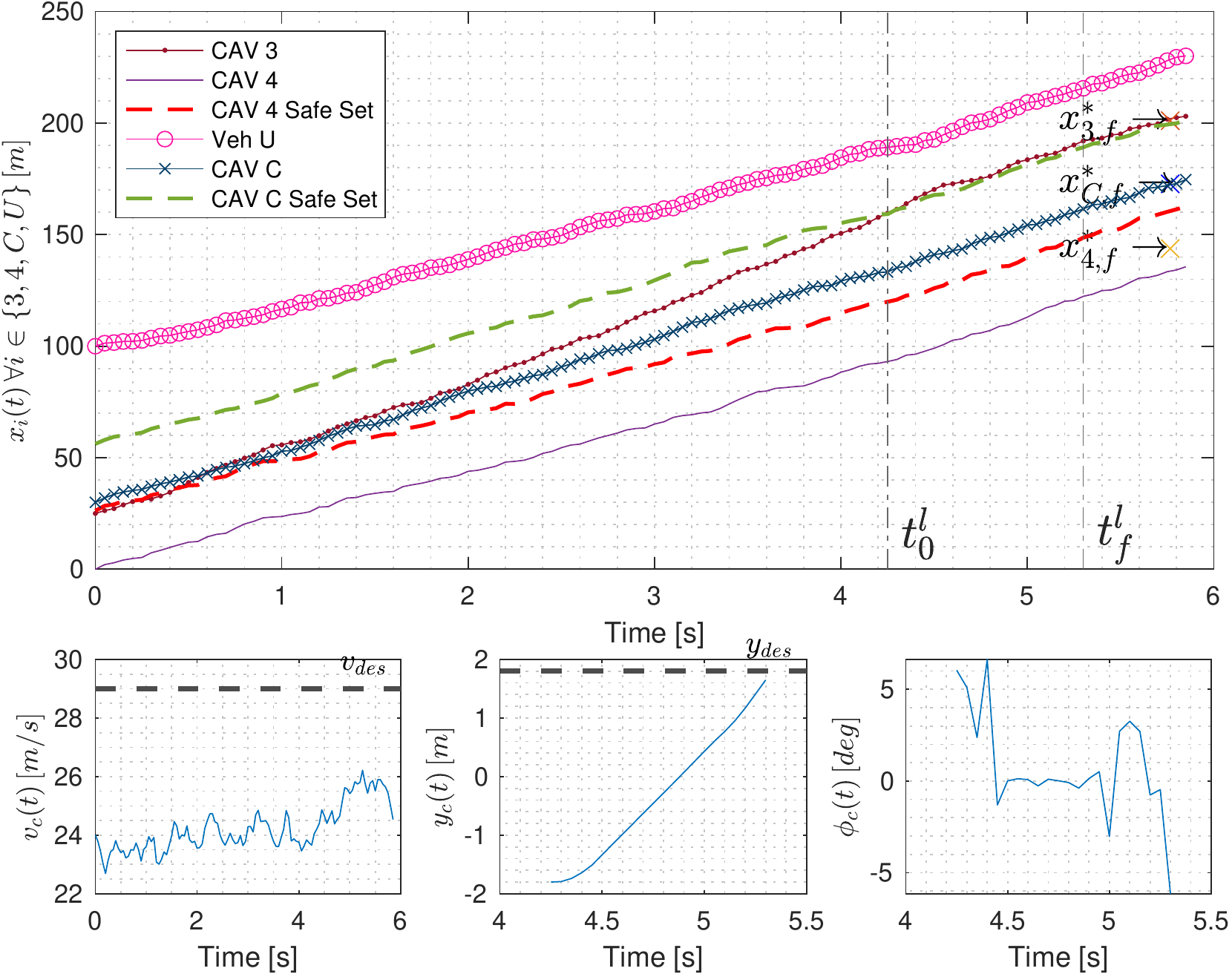}  
      \caption{\centering{CAV $C$ State history for Scenario 1}}
      \label{sub_fig:nominal_cav_C_states}
    \end{subfigure}   
    \begin{subfigure}{1\linewidth}
    \centering 
      \includegraphics[width=\linewidth, height=5cm]{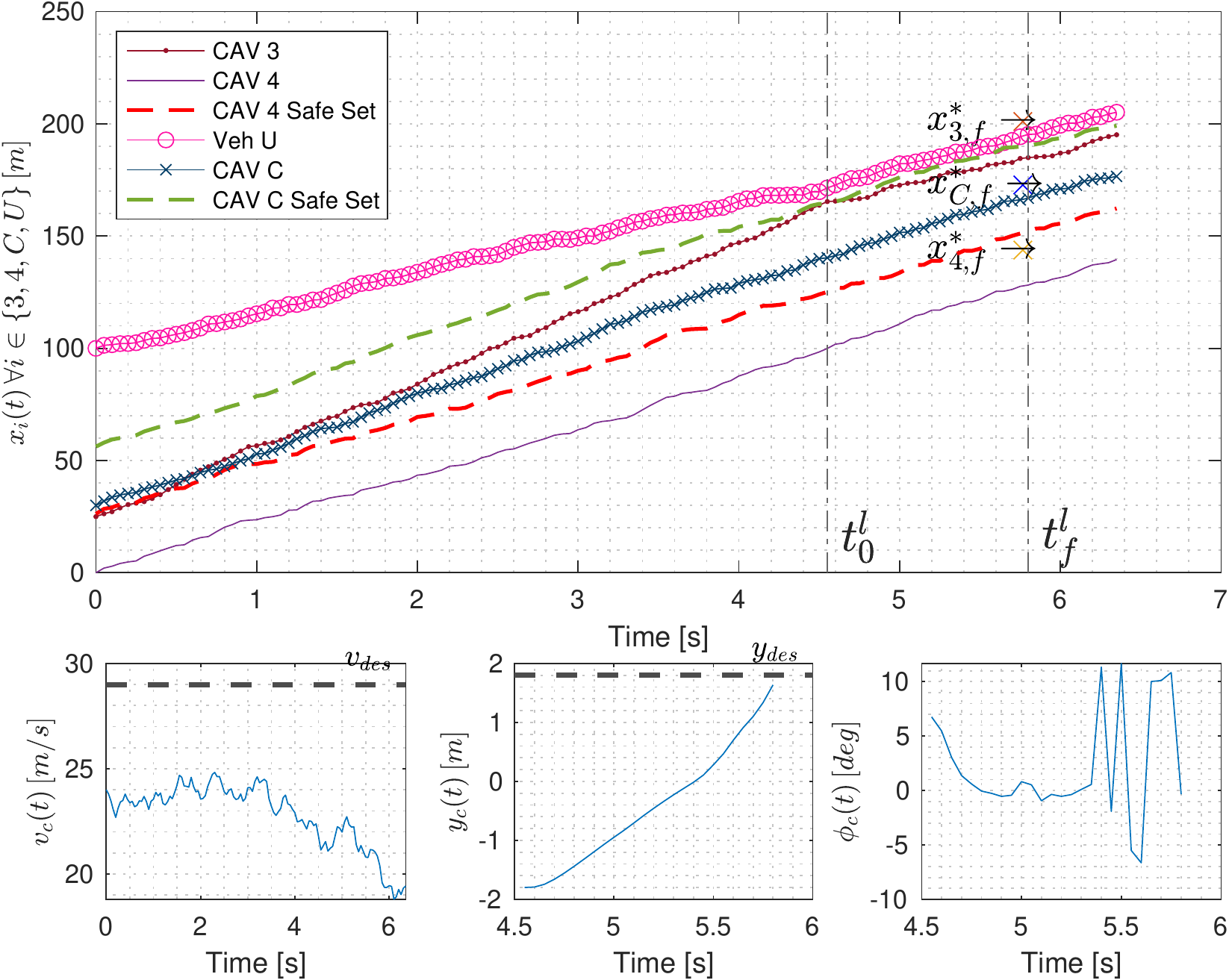}  
      \caption{\centering{CAV $C$ State history for Scenario 2}}
      \label{sub_fig:uncooperative_cav_C_states}
    \end{subfigure}\\
    \caption{\centering CAV $C$ sample state history and optimal merging triplet position history }
\end{figure}

In Table \ref{tab:energy_comparison}, it is shown that the CBF method has a longer maneuver time compared to the FxT-OCBF and OCBF methods. This suggests that the OCBF method ensures maneuver feasibility, while the fixed-time convergence helps with achieving convergence and synchronization. The CBF method exhibits lower acceleration peaks at the beginning of the maneuver since none of the safety constraints are activated. In contrast, the OCBF and FxT-OCBF methods experience higher control input deviations as they require more effort to closely track their reference inputs at the beginning. Nevertheless, in scenario 1, all three methods successfully perform the lane-changing maneuver. We illustrate the states for the lane-changing maneuver using the FxT-OCBF method in Fig. \ref{sub_fig:nominal_cav_C_states}, which depicts the positions of CAVs 3, 4, C, and the uncooperative vehicle U. Two vertical lines represent $t_0^l$ and $t_f^l$, indicating the times at which the lateral maneuver took place.

We will now analyze the second scenario, where vehicles U and F experience constant deceleration. In this case, as shown in Fig. \ref{sub_fig:diff_ocp_2}, only the FxT-OCBF and OCBF methods were able to successfully perform the lateral maneuver. However, for the CBF method, we included the trajectory history until the maneuver became infeasible. It can be observed that the FxT-OCBF method requires lower acceleration corrections compared to the OCBF method. This demonstrates the effectiveness of the spatio-temporal constraint in providing smoother and shorter trajectories when considering the merging constraint.

Due to the deceleration of vehicle U, the speed difference of the FxT-OCBF method is significantly lower compared to the other methods. This is because the maneuver takes place in shorter times, avoiding the activation of the rear-end safety constraints between CAV C and vehicle U, as well as between CAV 1 and vehicle F. These observations are evident in Fig. \ref{sub_fig:uncooperative_cav_C_states}, where the rear-end safety constraint of CAV C with respect to vehicle U is only activated during the execution of the lateral maneuver. Lastly, it can be noticed in Fig. \ref{sub_fig:uncooperative_cav_C_states} that the safety constraint is never violated at any instance, demonstrating the robustness of the proposed controller against additive disturbances.

Given Table \ref{tab:disruption_results} and Table \ref{tab:energy_comparison}, we can observe that for all three methods, the average energy consumption is higher than for the ideal scenario representing the unconstrained OCP solution. However, in both scenarios, the FxT-OCBF method achieves lower energy consumption and disruption values compared to the other methods, with orders of magnitude smaller disruption than the CBF case. The OCBF method shows similar energy consumption values to the FxT-OCBF method in scenario 1, which can be attributed to the feedback law provided by the reference states, helping with the feasibility of the maneuver while staying close to the optimal control inputs. In scenario 2, the fixed-time convergence allows for faster maneuver execution times, resulting in feasible and efficient maneuvers.


\begin{table}[ptb]
\vspace*{-0.5\baselineskip}
\centering
\caption{Average performance metric comparison results for scenarios 1 and 2 }
\label{tab:disruption_results}
\resizebox{\columnwidth}{!}{%
\begin{tabular}{|c|c|c|c|}
\hline
{\ul \textbf{Scenario}} &
  {\ul \textbf{Method}} &
  \textbf{$\mathbf{\frac{1}{N}\sum \limits_{i=1}^N{D}_i^v(t_0^l)\;[m^2/s^2]}$} &
  \textbf{$\mathbf{\frac{1}{N}\sum\limits_{i=1}^N{D}_i^x(t_0^l)\;[m^2]}$} \\ \hline
\rowcolor[HTML]{FFFC9E} 
1 & FxT-OCBF & 285.08   & 11.79  \\ \hline
1 & OCBF           & 472.07   & 15.63  \\ \hline
1 & CBF            & 2730.36  & 128.14 \\ \hline
\rowcolor[HTML]{FFFC9E} 
2 & FxT-OCBF & 232.03   & 43.52  \\ \hline
2 & OCBF           & 240.13   & 44.33  \\ \hline
2 & CBF            & 18299.85 & 692.16 \\ \hline
\end{tabular}%
}
\end{table}
\begin{table}[tbp]
\centering
\caption{Average Energy Comparison}
\label{tab:energy_comparison}
\resizebox{\columnwidth}{!}{%
\begin{tabular}{|c|c|c|c|ccc|}
\hline
 &
   &
  \multicolumn{1}{l|}{} &
  \multicolumn{1}{l|}{} &
  \multicolumn{3}{c|}{\textbf{$\mathbf{\frac{1}{N}\sum\limits_{i=1}^N\left(\int_{t_0}^{t_f} \frac{1}{2} u(t) d t\right)}$}} \\ \cline{5-7} 
\multirow{-2}{*}{{\ul \textbf{Scenario}}} &
  \multirow{-2}{*}{{\ul \textbf{Method}}} &
  \multicolumn{1}{l|}{\multirow{-2}{*}{\textbf{$\mathbf{t_f^*\, [s]}$}}} &
  \multicolumn{1}{l|}{\multirow{-2}{*}{\textbf{$\mathbf{t_0^l\, [s]}$}}} &
  \multicolumn{1}{c|}{\textit{Ideal}} &
  \multicolumn{1}{c|}{\textit{Actual}} &
  \textit{\% Diff} \\ \hline
\rowcolor[HTML]{FFFC9E} 
1 &
  FxT-OCBF &
  5.77 &
  4.25 &
  \multicolumn{1}{c|}{\cellcolor[HTML]{FFFC9E}11.93} &
  \multicolumn{1}{c|}{\cellcolor[HTML]{FFFC9E}23.43} &
  {\color[HTML]{FE0000} 96.45} \\ \hline
1 &
  OCBF &
  5.77 &
  5.95 &
  \multicolumn{1}{c|}{11.93} &
  \multicolumn{1}{c|}{25.32} &
  112.30 \\ \hline
1 &
  CBF &
  5.77 &
  13.75 &
  \multicolumn{1}{c|}{11.93} &
  \multicolumn{1}{c|}{27.68} &
  132.06 \\ \hline
\rowcolor[HTML]{FFFC9E} 
2 &
  FxT-OCBF &
  5.77 &
  4.55 &
  \multicolumn{1}{c|}{\cellcolor[HTML]{FFFC9E}11.93} &
  \multicolumn{1}{c|}{\cellcolor[HTML]{FFFC9E}20.94} &
  {\color[HTML]{FE0000} 75.51} \\ \hline
2 &
  OCBF &
  5.77 &
  6.2 &
  \multicolumn{1}{c|}{11.93} &
  \multicolumn{1}{c|}{37.23} &
  212.13 \\ \hline
2 &
  CBF &
  5.77 &
  Unfeasible &
  \multicolumn{1}{c|}{11.93} &
  \multicolumn{1}{c|}{39.61} &
  232.10 \\ \hline
\end{tabular}%
}
\vspace*{-\baselineskip}
\end{table}

%% file: sections/Conclusions.tex
We developed a decentralized optimal control framework for multiple CAVs that is robust against disturbances and uncooperative vehicle behavior by utilizing control barrier functions and spatio-temporal constraints for this purpose. The proposed controller employs an unconstrained analytical solution as a feedback reference controller, which is tracked using the proposed controller with fixed-time convergence guarantees.
Simulation results demonstrate the effectiveness of our controller in executing cooperative lane-changing maneuvers efficiently, even in the presence of disturbances and uncooperative vehicles that limit the feasibility of the maneuver. The simulations show the time convergence guarantees while minimizing energy consumption and disruption when compared to two other methods that neglect these factors.

Future work will focus on providing comfort guarantees during maneuver execution. Additionally, we plan on exploring different levels of cooperation while allowing human-driven vehicles to implicitly cooperate with CAVs through rule-compliance methods.